\documentclass[fleqn,usenatbib]{mn2e}
\usepackage{times}
\usepackage{amsmath}
\usepackage{graphicx}
\usepackage{booktabs} 
\usepackage{dcolumn}

\def\teff{{\rm T}$_{\rm eff}$}
\def\logteff{$\log$(${\rm T}_{\rm eff}$)}
\def\logg{$\log g$}
\def\feh{[Fe/H]}
\def\av{A$_{\rm V}$}
\def\ag{A$_{\rm G}$}
\def\mg{M$_{\rm G}$}
\def\rv{R$_{\rm V}$}
\def\aabun{[$\alpha$/Fe]}

\def\mr{$\epsilon_{\rm sys}$}
\def\absmr{$|\epsilon_{\rm sys}|$}
\def\mar{$\epsilon_{\rm mae}$}
\def\rms{$\epsilon_{\rm rms}$}

\def\chisq{$\chi^2$}
\def\sens{\mathbfss S}

\def\ltsim{\:{_<\atop{^\sim}}\:}

\def\ilium{{\sc ilium}}

\title[Parameter estimation with \ilium]{The ILIUM forward modelling algorithm for multivariate parameter estimation and its application to derive stellar parameters from Gaia spectrophotometry}

\author[C.A.L.\ Bailer-Jones]
{C.A.L.\ Bailer-Jones\\
Max-Planck-Institut f\"ur Astronomie, K\"onigstuhl 17, 69117
Heidelberg, Germany\thanks{Email: calj@mpia.de}\\
}

\begin{document}

\date{Accepted 2009 November 26.  Received 2009 November 26; in original form 2009 October 19}

\maketitle

\label{firstpage}

\begin{abstract} 
  I introduce an algorithm for estimating parameters from multidimensional data based on forward modelling.  It performs an iterative local search to effectively achieve a nonlinear interpolation of a template grid.  In contrast to many machine learning approaches it avoids fitting an inverse model and the problems associated with this.  The algorithm makes explicit use of the sensitivities of the data to the parameters, with the goal of better treating parameters which only have a weak impact on the data. The forward modelling approach provides uncertainty (full covariance) estimates in the predicted parameters as well as a goodness-of-fit for observations, thus providing a simple means of identifying outliers. I demonstrate the algorithm, \ilium, with the estimation of stellar astrophysical parameters (APs) from simulations of the low resolution spectrophotometry to be obtained by Gaia. The AP accuracy is competitive with that obtained by a support vector machine. For zero extinction stars covering a wide range of metallicity, surface gravity and temperature, \ilium\ can estimate \teff\ to an accuracy of 0.3\% at G=15 and to 4\% for (lower signal-to-noise ratio) spectra at G=20, the Gaia limiting magnitude (mean absolute errors are quoted). \feh\ and \logg\ can be estimated to accuracies of 0.1--0.4\,dex for stars with G\,$\leq18.5$, depending on the magnitude and what priors we can place on the APs.  If extinction varies a priori over a wide range (0--10\,mag) -- which will be the case with Gaia because it is an all sky survey -- then \logg\ and \feh\ can still be estimated to 0.3 and 0.5\,dex respectively at G=15, but much poorer at G=18.5. \teff\ and \av\ can be estimated quite accurately (3--4\% and 0.1--0.2\,mag respectively at G=15), but there is a strong and ubiquitous degeneracy in these parameters which limits our ability to estimate either accurately at faint magnitudes. Using the forward model we can map these degeneracies (in advance), and thus provide a complete probability distribution over solutions.  Additional information from the Gaia parallaxes, other surveys or suitable priors should help reduce these degeneracies.  \end{abstract}


\section{Introduction \label{introduction}}

Inferring parameters from multidimensional data is a common task in astronomy, whether this be inference of cosmological parameters from CMB experiments, photometric-redshifts of galaxies or physical properites from stellar spectra.  Important questions about the structure and evolution of stars and stellar populations require knowledge of abundances and ages, which must be obtained spectroscopically via the stellar atmospheric parameters effective temperature (\teff), surface gravity (\logg) and iron-peak metallicity (\feh).

Numerous publications have presented methods for estimating such astrophysical parameters (APs) from spectra. The methods can be divided into two broad categories. The first, based on high resolution and high signal-to-noise ratio (SNR) spectra, makes use of specific line indices selected to be sensitive predominantly to the phenomena of interest.  Examples include the detection of BHB stars via Calcium and Balmer lines (e.g.\ Brown~\citealp{brown03}) and spectral type classification of M, L and T dwarfs via molecular band indices (e.g.\ Hawley et al.~\citealp{hawley02}). In each case these methods use a specific, identified phenomenon and soare generally limited to a narrow part of the AP space where the values of the other APs are reasonably well known. While simple and useful, these methods do not use all available information nor can they normally be used with low resolution data, because the effects of individual APs cannot then be separated.

The second category of methods is global pattern recognition, which try to use the full set of available data. These are generally used when we want to estimate one or more APs over a wide range of parameter space.  In such cases there is rarely a simple relation between a single feature and the physical quantity of interest. (An example is determination of surface gravity, where the relevant lines to use change with temperature.) We must therefore infer which features are relevant to which APs in which parts of the AP space. As this is generally a nonlinear, multidimensional problem, the standard approach is to use a machine learning algorithm to learn the mapping from the data space to the AP space based on labelled template spectra (spectra with known APs).  Various models have been used in astronomy for a variety of problems, including (to name just a few): neural networks for stellar parameter estimation (e.g.\ Re Fiorentin et al.~\citealp{rf07}) or photometric redshift estimation (e.g.\ Firth et al.~\citealp{firth03}); support vector machines for quasar classification (e.g.\ Gao et al.~\citealp{gao08}, Bailer-Jones et al.~\citealp{cbj08}) or galaxy morphology classification (e.g.\ Huertas-Company et al.~\citealp{hc08}); classification trees for identifying cosmic rays in images (e.g.\ Salzberg et al.~\citealp{salzberg95}); linear basis function projection methods (e.g.\ the method of Recio-Blanco et al.~\citealp{recioblanco06} for spectral parameter estimation, which constructs the basis functions from model spectra).  More examples of machine learning methods and their use in astronomy can be found in the volume edited by Bailer-Jones~\citep{cbj08b}.

Note that the first category of methods, line indices, is really just a special case of the second in which drastic feature selection has taken place to enable use of low (one or two) dimensional models. In both cases we must learn some relationship ${\rm AP} = g'({\rm Data})$ from a  model or based on some labelled templates.
But this is an {\em inverse} relation: more than one set of APs may fit a given set of data (e.g.\ a low extinction cool star or high extinction hot star could produce the same colours). Despite this non-uniqueness we nonetheless try and fit a unique model.
This causes fitting problems which become more severe the lower the quality of the data (the lower the number of independent measures) and the larger the number of APs we want to estimate from it, and could lead to poor AP estimates or biases.
In contrast, the forward mapping (or {\em generative model}), ${\rm Data} = g({\rm AP})$ is unique, because this is a causal, physical model (e.g.\ a stellar atmosphere and radiative transfer model of a spectrum).
A further issue is that the model must learn the sensitivity of each input to each AP (and how noise affects this). Yet this information we in principle have already from the gradients of the generative model.

A pattern recognition method which tries to overcome some of these problems is $k$-nearest neighbours, which has also been applied to many problems in astronomy (e.g.\ Katz et al.~\citealp{katz98}, Ball et al.~\citealp{ball08}; plus extensions thereof such as kernel density estimation, e.g.\ Richards et al.~\citealp{richards04}, or the
method of Shkedy et al.~\citealp{shkedy07}, which converts the distances into likelihoods and uses priors to create a
full probabilistic solution.) In many ways this is the most natural way to solve the problem: we create a grid of labelled templates and find which are closest to our observation (perhaps smoothing over several neighbours). On the assumption that the APs vary smoothly with the data between the grid points, this may provide a good estimate. But for this to be accurate (and not too biased), the grid must be sufficiently dense that 
multiple grid points lie within the error ellipse of the observation (the covariance of these neighbours then provides a measure of the uncertainty in the estimated APs). If the SNR is high, the grid must therefore be very dense.
Moreover, as the number of APs increases, the required grid density grows exponentially with it: 
For a stellar parametrization problem with 5 APs we might need an average of 100 samples per AP, resulting in $100^{5}=10^{10}$ templates.
There is also the issue of what distance metric to use. The covariance-weighted Mahalanobis distance is often used, but it ignores the sensitivities of the inputs. (If some inputs are sometimes dominated by irrelevant cosmic scatter, this will add unmodelled noise to the distance estimate.) 
This is a problem when we have a mix of APs, some of which have a large and others a small impact on the variance (``strong'' and ``weak'' APs).
If we just use the Mahalanobis distance we loose sensitivity to the weak APs.

We could overcome these problems if we did on-the-fly interpolation of the template grid to generate new templates
as we need them. Running stellar models is far too time consuming for this, but also unnecessary because the generative model is smooth: We can instead fit a forward model to a low density grid of templates as an approximation to the generative model.
As we shall see, possession of a forward model opens up opportunities not available to the inverse methods, such as direct uncertainty estimates and goodness-of-fit assessment of the solution.

In this article I introduce an algorithm for AP estimation based on this forward modelling idea and iterative interpolation.  I will demonstrate it using simulations of low resolution spectra to be obtained from the Gaia mission (e.g.\ Lindegren et al.~\citealp{lindegren08}).\footnote{\tt http://www.rssd.esa.int/Gaia} Gaia will observe more than $10^9$ stars down to 20$^{th}$ magnitude over the whole Galaxy, stars which a priori span a very wide range in several APs. This includes the line-of-sight extinction parameter, \av, which must be estimated accurately if we want to derive intrinsic stellar luminosities from the Gaia parallaxes.  AP estimation (Bailer-Jones~\citealp{cbj05}) is therefore an integral part of the overall Gaia data processing and comprises one of the Coordination Units in the Gaia Data Processing and Analysis Consortium (DPAC)
(Mignard \& Drimmel~\citealp{aoresponse}, O'Mullane et al.~\citealp{omullane07}).

I will now describe the basic algorithm (section~\ref{algorithm}). In section~\ref{data} I then introduce the simulated Gaia spectroscopy to which \ilium\ is applied, with the results and discussion thereof presented in sections \ref{sect:tefflogg}, \ref{sect:tefffeh} and~\ref{sect:2d1d}. The latter section also reports on a strong and ubiquitous degeneracy between \teff\ and \av.  I summarize and conclude in section~\ref{conclusions}.  Additional plots, results and discussions can be found in a series of four Gaia technical notes (Bailer-Jones 2009a,b,c,d) available from {\tt http://www.mpia.de/Gaia}.

\section{The \ilium\ algorithm} \label{algorithm}

I outline the algorithm using the terminology of spectra and stellar astrophysical parameters, although it is quite general and applies to any multivariate data.  Table~\ref{tab:notation} summarizes the notation.  ``Band'' refers to a flux measurement in the spectrum. In general it could be a photometric band, a single pixel in the spectrum, or a function of many pixels.

\begin{table}
\begin{center}
\caption{Notation\label{tab:notation}}
\begin{tabular}{ll}
\hline
$I$         & number of bands (pixels in spectrum)\\
$i$         & counter over band, $i=1 \dots I$ \\
$p_i$       & photon counts in band $i$ ($\bmath p$ is a spectrum)\\
$J$         & number of APs (astrophysical parameters)\\
$j$         & counter over AP, $j=1 \dots J$ \\
$\phi_j$    & AP $j$ ($\bmath \phi$ is a set of APs)\\
$s_{ij}$    & sensitivity of band $i$ to AP $j$, $\frac{\partial p_i}{\partial \phi_j}$ \\
\sens         & sensitivity matrix, $I \times J$ matrix with elements $s_{ij}$ \\
$f_i(\bmath \phi)$ & forward model for band $i$ \\
$n$         & iteration; e.g.\ $\phi(n)$ is the AP at iteration $n$ \\
\hline
\end{tabular}
\end{center}
\end{table}

\subsection{Forward modelling}

I will call the true relationship between the APs and the flux in a band $i$ the {\em generative
  model}, $g_i(\bmath \phi)$. This provides the observed spectrum for a given set of APs and thus encapsulates the underlying stellar model, radiative transfer, interstellar extinction, instrument model etc. (There is a separate function for each band, but for simplicity I will refer to it in the singular.) Generally we don't have an explicit function for this model, so it remains unknown. All we have for doing AP estimation is a discrete {\em template grid} of example spectra with known APs generated by the generative model.  We approximate the generative model using a {\em forward model}, $f_i(\bmath \phi)$, which is a (nonlinear) parametrized fit to this grid and provides flux estimates at arbitrary APs, i.e.\ off the grid. (Forward models can be fit independently for each band.) Demanding the forward models to be continuous functions ensures we can also use them to calculate the sensitivities, which by definition are the gradients of the flux with respect to each AP. The forward model fitting is done just once for a given grid and is kept fixed when predicting APs. In other words it is a training procedure.

\subsection{Core algorithm}

The basic idea of \ilium\ is to use the Newton-Raphson method to find that forward model-predicted spectrum (and associated APs) which best fits the oberved spectrum.


\begin{figure}
\begin{center}
\includegraphics[width=0.31\textwidth, angle=-90]{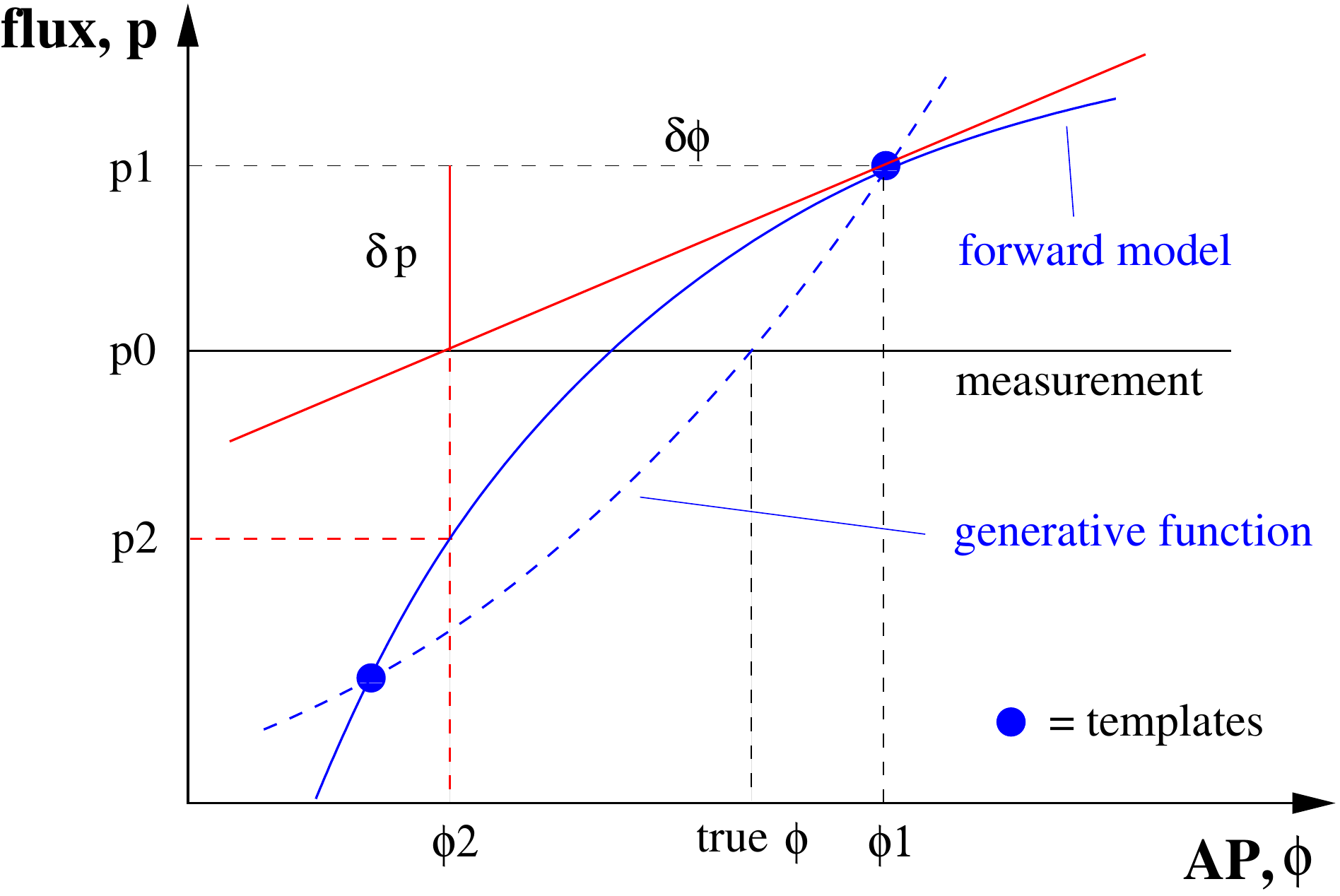}
\caption{Sketch of the search method for one band ($I$=1) and one AP ($J$=1). The dashed blue curve is the (unknown) generative model, and the solid blue curve is the forward model (our approximation to the generative model) formed
  by fitting a function to the templates. (The difference between the two is exaggerated.) The straight red line
  shows the local linear approximation of the forward model (tangent at $\phi_1$) used to
  calculate the first AP step. (In the case show the
  forward model could be inverted. But this is not generally
  the case, not even in one dimension if it has a turning
  point.)} \label{fig:principle}
\end{center}
\end{figure}

In detail, the algorithm is as follows (Fig.~\ref{fig:principle}). Consider first a single AP and single band. The measurement is $p(0)$ and we want to estimate its AP.
The forward model, $\hat p = f(\phi)$, has been fit and remains fixed. The procedure is as follows ($n$ is the iteration number) 
\begin{enumerate}
\item Initialize: find nearest grid neighbour to $p(0)$, i.e.\ the one which minimizes the sum-of-squares residual $\delta \bmath p^T \delta \bmath p$.  Call this $[p(1), \phi(1)]$. $\phi(1)$ is the initial AP estimate.
\item Use the forward model to calculate the local sensitivities, $\frac{\partial p}{\partial \phi}$, at the current AP estimate.
\item Calculate the discrepancy (residual) between the predicted flux and the measured flux, $\delta p(n) = p(n) - p(0)$. 
\item Estimate the AP offset as \, $\delta \phi(n) = \left (\frac{\partial \phi}{\partial p} \right)_{\phi(n)} \times \delta p(n) $, i.e.\ a Taylor expansion truncated to the linear term. (Note that this partial derivative is the reciprocal sensitivity.)  
\item Make a step in AP space, $ \phi(n+1) = \phi(n) - \delta \phi(n)$, {\em toward} the better estimate. This is the new AP prediction.
\item Use the forward model to predict the corresponding (off-grid) flux, $p(n+1)$
\item Iterate steps ii--vi until convergence is achieved or a stop is imposed.
\end{enumerate}
The algorithm is basically minimizing $|\delta \bmath p|$.  At each iteration we obtain an estimate of the APs (step v) and the corresponding spectrum (step vi). Convergence could be defined in several ways, e.g.\ when changes in the spectrum or the APs (or their rate of change) drop below some threshold.  Alternatively we could simply stop after some fixed number of iterations.  There is no guarantee of convergence.  For example, if the AP steps were sufficiently large to move to a part of the function with a sensitivity of the opposite sign, then the model could diverge or get stuck in a limit cycle.  Likewise, if initialized too far from the true solution the algorithm could become stuck in a local minimum far from the true solution.  For this, and other reasons, the algorithm in practice has some additional features (discussed in section~\ref{sect:practical}). Also,

\subsection{Generalization to multiple APs and bands}

In general we have several bands and several APs. The flux perturbation 
due to small changes in the APs is then
\begin{equation}
\delta {\bmath p} \, = \, \sens \, \delta{\bmath \phi}
\label{eqn:deltap}
\end{equation}
where $\sens$ is the $I \times J$ sensitivity matrix with elements $s_{ij} = {\partial p_i}/{\partial \phi_j}$.
Note that $I>J$. Multiplying this equation on the left
by $(\sens^T \sens)^{-1}\sens^T$ gives
\begin{equation}
\delta {\bmath \phi} \, = \, (\sens^T \sens)^{-1}\sens^T \delta{\bmath p}
\label{eqn:deltaap}
\end{equation}
so the AP update equation (step v in the algorithm) becomes
\begin{equation}
{\bmath \phi}(n+1) = {\bmath \phi}(n) - (\sens^T \sens)^{-1}\sens^T \delta{\bmath p}(n)
\label{eqn:apupdate}
\end{equation}
The $I$ forward models are now functions of $J$ variables, and this turns out to be a critical matter.

\subsection{The forward model}\label{sect:forward_model}

The core algorithm just described can make use of any form for the forward model, on the condition that it provides values of the function and its first derivatives for arbitrary values of the APs.

The most obvious forward model would be a multidimensional, nonlinear regression of the form $\hat{p} = f({\bmath \phi})$, which in principle works for any number of APs. However, I found that it was difficult to get a model which simultaneously fits both \teff, a strong AP, and \logg, a weak AP to sufficient accuracy.  {\em Strong} here means that it explains much of the variance in the flux data, i.e.\ is a strong predictor of the flux. {\em Weak} is a relative term, indicating that the AP explains much less of the variance.  The reason for poor fits over the weak APs is that the model is fit by minimizing a single objective function, namely the error in reproducing the flux. As the weak AP has very little impact on the flux, its influence has little impact on the error, so the model optimization does little to produce a good fit over this AP.

\begin{figure}
\vspace*{-2ex}
\begin{center}
\includegraphics[width=0.4\textwidth]{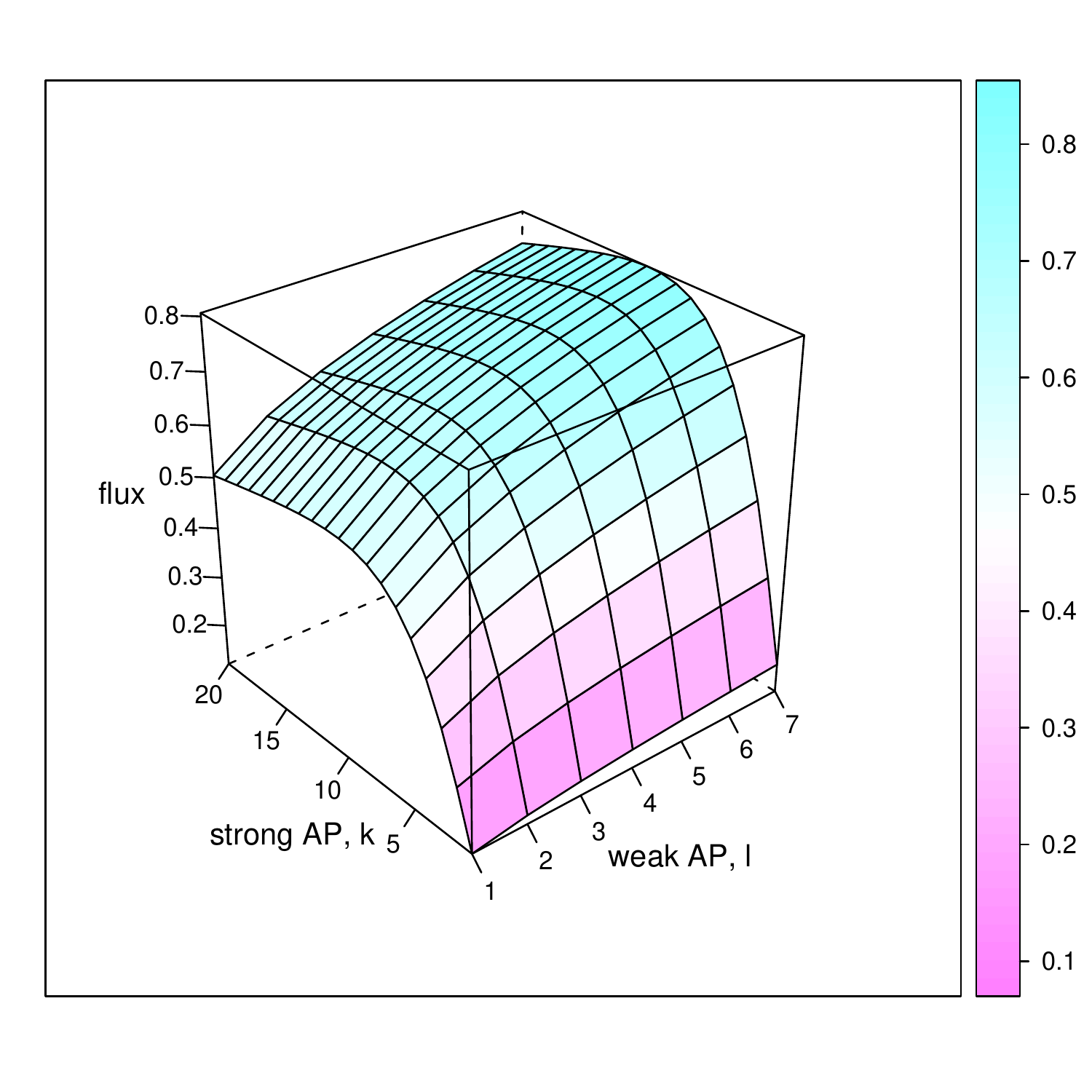}
\vspace*{-4ex}
\caption{Schematic diagram of a two-component forward model. Note the stronger variation in the flux in the direction of the strong AP (the contrast is typically much larger for the problems considered in this article).}\label{fig:forward_model_schematic}
\end{center}
\end{figure}

To overcome this problem I use a two-component forward model to separately fit the strong and weak APs. Consider the case of a single strong AP and a single weak AP (this is generalized later). The strong component is a 1D nonlinear function of the strong AP which is fit by marginalizing over the weak AP.  This fits most of the flux variation. Then, at each discrete value of the strong AP in the grid, we fit the residual flux as a function of the weak AP; these are the second components (also 1D).  They provide a flux increment dependent on the weak AP, which is added to the flux predicted given the strong AP.  A schematic illustration of such a two-component forward model is illustrated in Fig.~\ref{fig:forward_model_schematic}.

\begin{figure} \begin{center}
    \includegraphics[width=0.33\textwidth]{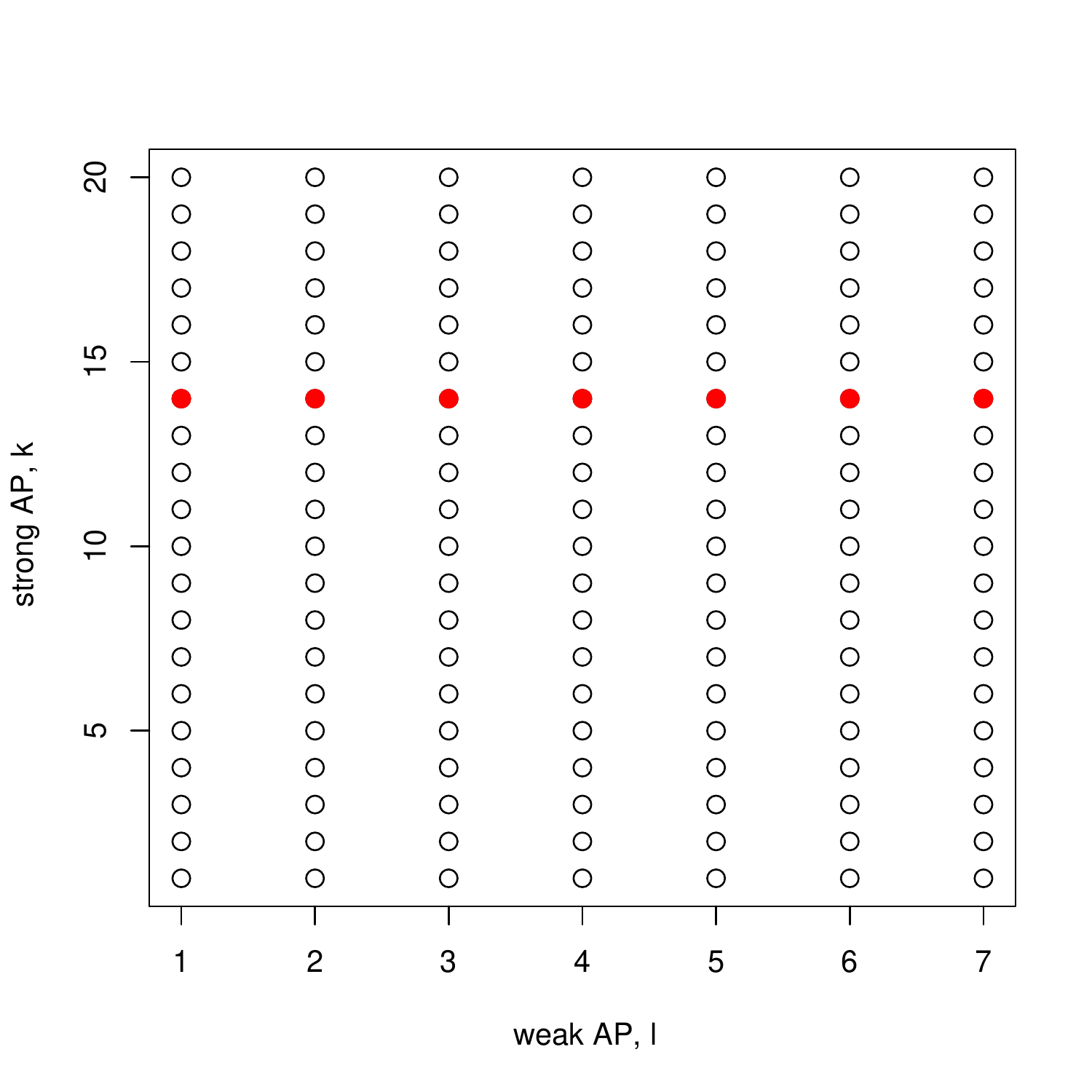} \caption{Schematic diagram
      of the grid of AP values, $k = 1 \ldots 20$, $l = 1 \dots 7$. The solid (red) points denote those used to
      fit the weak component of the forward model for one value of
      the strong AP.}\label{fig:schematic_grid}
\end{center}
\end{figure}

\begin{figure} 
\begin{center}
    \includegraphics[width=0.33\textwidth]{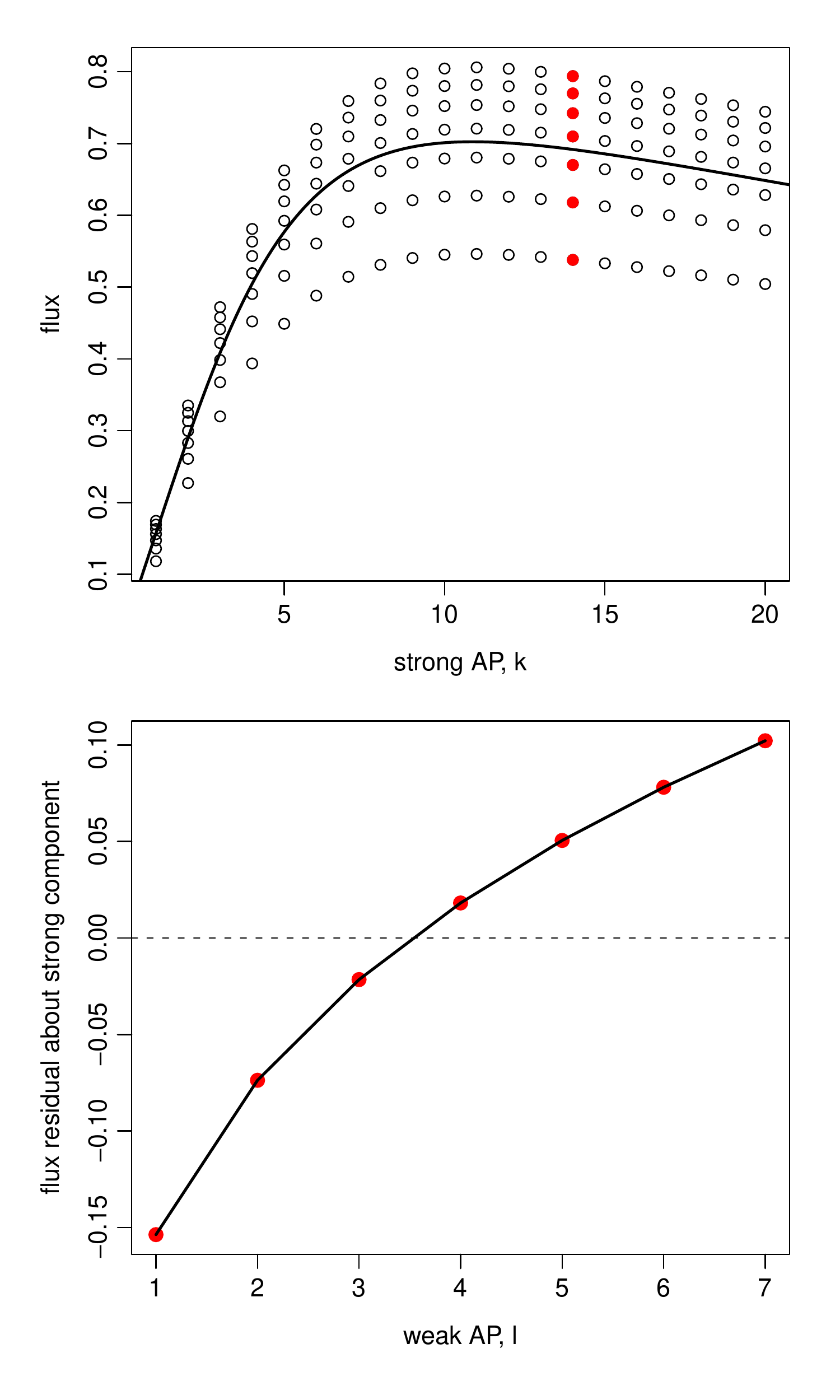} 
    \caption{Schematic illustration of the two components of the
      forward model fit over all the strong and weak points in the grid in Fig.~\ref{fig:schematic_grid} Top: the fit over the strong APs. Bottom: one of
      the fits over the weak APs (the solid/red points in the top panel at
      k=14).}\label{fig:two_component_forward_model}
\end{center}
\end{figure}

To be more precise the model is fit as follows. Let $\phi^S$ denote a strong AP, $\phi^W$ a weak AP and $f_i(\phi^S, \phi^W)$ the complete (2D) forward model for band $i$.  Let subscripts $k$ and $l$ denote specific values in the grid of the strong and weak APs respectively (see Fig.~\ref{fig:schematic_grid}).  We model the flux at an arbitrary AP point as
\begin{equation}
f_i(\phi^S, \phi^W) \, = \, f^S_i(\phi^S) \, + \, f^W_{i,k}(\phi^W ; \phi^S=\phi^S_k) 
\end{equation}
where both $f^S_i$ and $ f^W_{i,k}$ are 1D functions. $f^S_i(\phi^S)$ is the single strong component (for band $i$). It 
is a fit 
to the average value of $\phi^W$ at each $\phi^S$ (the curve in the top panel of Fig.~\ref{fig:two_component_forward_model}). 
$f^W_{i,k}$ is the $k^{th}$ weak component, which is a fit with respect to the weak AP with the strong AP fixed at $\phi^S_k$ (e.g.\ the solid/red points in Fig.~\ref{fig:schematic_grid}), i.e.\ it is fit to the residuals
\[
\{ \, p(\phi^S = \phi^S_k) - \overline{ p(\phi^S = \phi^S_{k^{'}}) } \, \}
\]
as illustrated in in the bottom panel of Fig.~\ref{fig:two_component_forward_model}.

This fitting approach clearly requires us to have a ``semi-regular'' grid: one which has a range of values of the weak AP for each value of the strong AP. (This requirement is easily fulfilled when using synthetic grids.) The number of weak components in the model is equal to the number of unique values of $\phi^S$, 20 in this schematic case.
If we have more than one strong or weak component then we raise the dimensionality of the strong or weak component (see section~\ref{sect:2d1d}).

Applying the forward model is easy. Given $(\phi^S, \phi^W)$ we
\begin{enumerate}
\item evaluate $f^S_i$, the strong component;
\item find the nearest value, $\phi^S_k$, in the grid to $\phi^S$, i.e.\ identify the closest weak component;
\item evaluate $f^W_{i,k}$, the increment from the weak component;
\item sum the two components, $f^S_i + f^W_{i,k}$, to give the forward model prediction.
\end{enumerate}
Although the weak component we use changes discontinuously as $\phi^S$ varies, the weak component is only specifying
an {\em increment} to the strong component fit.  
As both components are smooth in their respective APs the combined function is also smooth along any direction parallel to the AP axes. It is not smooth along arbitrary directions, but this is unimportant because explicit calculations with the forward model (e.g.\ of the sensitivities) are only carried out parallel to the AP axes.

\subsection{Practical algorithm}\label{sect:practical}

The practical realities of working with real (noisy) data mean that the basic algorithm should be extended in order to make it more robust. These I now describe together with other implementation aspects used for the experiments described later.  For the purposes of this paper the algorithm has been implemented in R\footnote{{\tt http://www.r-project.org}}.  A Java implementation is in progress, which will be necessary for larger scale applications.

\subsubsection{Standardized variables}

The spectral variables are observed photon counts (to within an irrelevant constant factor).
The APs are all on logarithmic scales: \av\ in magnitudes, \logg\ and \feh\ in dex, and \logteff.  
In order to bring each variable to the same level I
standardize the flux in each band and each AP (linearly scale each to have zero mean and unit variance).
If there were correlations in the spectra these could be removed by ``sphereing'' (``prewhitening''), a covariate generalization of standardization which gives the data unit diagonal covariance (e.g.\ Bishop~\citealp{bishop06}).

\subsubsection{Forward model functions}\label{fmfunctions}

The strong and weak components of the forward model are fit using smoothing splines (e.g.\ Hastie et al.~\citealp{hastie01}). Conventional cubic splines have the drawback that one must control their complexity (smoothness) using the number and position of the knots. Smoothing splines circumvent this problem by setting a knot at every point (which would overfit the data) and then applying a smoothing penalty which is controlled by specifying the effective degrees-of-freedom (dof). I set this by trial and error via inspection of the resulting fits.  For the 2D problems TG (\teff\ and \logg) and TM (\teff\ and \feh) described in sections~\ref{sect:tefflogg} and~\ref{sect:tefffeh}, both the strong and the weak model splines are 1D.  The strong model uses dof\,$ = n_{\rm Teff}/2=16.5$ where $n_{\rm Teff}$ is the number of unique \teff\ points.  As the maximum number of \logg\ points is 10 (for the training data), and because the variation with \logg\ is quite smooth, I set the dof for these fits to be 4. However, many of the \teff\ values in the training grid have fewer \logg\ or \feh\ points: To avoid overfitting, if $n_{\rm logg}(T_{\rm eff}) \leq 4$ then a linear fit is used. If $n_{\rm
  logg}(T_{\rm eff}) = 1$, then no fit is performed and this weak component of the forward model is zero.
(Practical aspects of higher dimensional forward models are described in section~\ref{sect:2d1d}.)

The forward model is always fit to noise-free data. It is well known (and the author's experience) that inverse modelling methods such as ANNs and SVMs perform best when trained on data with a similar noise level as the target data (e.g.\ Snider et al.~\citealp{snider01}).This precipitates the need for multiple models when used on survey data with a range of SNRs, something which is not an issue for \ilium.


\subsubsection{Sensitivity estimation}

If the forward model is a simple analytic function then it may have
analytical first derivatives which can be used to calculate the
sensitivities. But in the general case we can use the method of first
differences
\begin{equation}
\left . \frac{\partial p}{\partial \phi_j} \right |_{\bmath \phi} \, \simeq \, \frac{f(\bmath \phi + \delta \phi_j) - f(\bmath \phi - \delta \phi_j)}{2 \, \delta \phi_j}
\label{eqn:firstdiff}
\end{equation}
I select $\delta \phi_j$ to be slightly smaller than the maximum precision a priori possible in an AP.  As the forward model must be smooth at this resolution, the first difference approximation is sufficiently accurate.  For the examples shown later, I choose $\delta \phi_j$ to be 0.05 dex for \logg\ and \feh, 0.0005 for \logteff\ (0.1\% for \teff) and 0.03\,mag for \av. 

\subsubsection{Lower limit on sensitivities (singularity avoidance)}

Given that the AP updates depend upon the inverse of the sensitivity matrix (equation~\ref{eqn:deltaap}), 
it is prudent to prevent the sensitivities being too small in order to avoid $\sens^T \sens$ being singular. For this reason, a lower limit is placed on the absolute value of each sensitivity, $s_{ij}$, of 0.001 (with $p$ and $\phi$ in standardized units). For the examples shown later, this limit rarely had to be applied in practice and had negligible impact on the results. 
To avoid singularity  $\sens^T \sens$ must also have a rank of at least $J$, i.e.\ to estimate $J$ APs we need at least $J$ independent measures in the spectrum.

\subsubsection{AP update contribution clipping}\label{sect:apcontclip}

Equation~\ref{eqn:deltaap} can be written
\begin{equation}
{\bmath \phi}(n+1) = {\bmath \phi}(n) - {\mathbfss M} \, \delta{\bmath p}(n)
\label{eqn:mupdate}
\end{equation}
where 
\begin{equation}
{\mathbfss M} = (\sens^T \sens)^{-1}\sens^T
\label{eqn:m}
\end{equation}
 is a $J \times I$ matrix. Equation~\ref{eqn:mupdate} gives
$J$ update equations, one for each AP. The update for AP $j$ can be written as the dot product of two vectors, the 
$j^{th}$ row of ${\mathbfss M}$, ${\bmath m}_j$ with ${\bmath p}(n)$, i.e.\
\begin{eqnarray}
\label{eqn:apcont}
\phi_j(n+1) &=& \phi_j(n) - {\bmath m}_j \, \delta{\bmath p}(n)  \nonumber \\
            &=& \phi_j(n) - \sum_i m_{ij} \delta p_i(n) \nonumber \\
            &=& \phi_j(n) - \sum_i u_{ij}(n)
\end{eqnarray}
which defines $u_{ij}$.  Thus we see that the update to AP $j$ is a sum over $I$
individual updates, which we can view as an update ``spectrum''. 
If we inspect these updates (see Bailer-Jones\citealp{cbj09a}) we see that
on occasion some are much larger than the others. 
This dominance of the update by just one of a few elements is undesirable, 
because they may be affected by noise
($\delta{\bmath p}(n)$ is a noisy measurement). For this reason, I clip
outliers in this spectrum. (It is valid to
compare the updates for different bands, because we work with
standardized fluxes.) To be robust, I set an upper (lower) limit which
is a multiple $c$ of the median of those points above (below) the
median.  Using the notation $\theta()$ to denote median, the limits
are
\begin{eqnarray}
u_{\rm upper} &=& \theta(u_i) + c[\theta(u_i > \theta(u_i)) - \theta(u_i)]  \nonumber \\
u_{\rm lower} &=& \theta(u_i) + c[\theta(u_i < \theta(u_i)) - \theta(u_i)]
\end{eqnarray}
I somewhat arbitrarily set $c=10$ so as to be relatively conservative in clipping.

\subsubsection{Upper limit on AP update step size}\label{sect:apupdates}

The AP steps at each iteration (equation~\ref{eqn:deltaap}) could be very large. This is undesirable, because the updates are based on a local {\em linear} approximation to the generative model. The code therefore imposes upper limits on the AP updates, corresponding to steps no larger than about 2.0\,dex in \logg\ and \feh, 0.04 in \logteff\ (10\% in \teff) and 0.3\,mag in \av.  A larger step size is permitted for the weaker APs because the initial nearest neighbour offset can be quite incorrect.  These limits are imposed more often for noisy data, but still relatively rarely. 

\subsubsection{Limit AP extrapolation}\label{sect:apextrapolate}

We do not expect the forward model to make good predictions beyond the
AP extremes of the grid, so I set upper and lower limits on the AP
estimates which \ilium\ can provide. These are set as $e$ times the
range of each AP, i.e.\
\begin{eqnarray}
{\rm upper~limit} &=& \max{\phi_j} + e(\max{\phi_j} - \min{\phi_j}) \nonumber \\
{\rm lower~limit} &=& \min{\phi_j} - e(\max{\phi_j} - \min{\phi_j})
\end{eqnarray}
I set $e=0.1$.

\subsubsection{Stopping criterion}

The algorithm is simply run for a fixed number of iterations (20). We often observe good natural convergence, so a more sophisticated stopping criterion is not applied at this time, although it may be important on high variance data sets (low SNR or more APs).

\subsection{Performance statistics}\label{errstat}

The model performance is assessed via the AP residuals (estimated minus true, $\delta \phi$) on an evaluation data set.
I report three statistics : (1) the root-mean-square (RMS) error,
which I abbreviate with \rms; (2) the mean absolute error, $\overline{|\delta \phi|}$, abbreviated as \mar; (3) the mean residual, $\overline{\delta \phi}$, a measure of the systematic error, abbreviated as \mr. 
(Note that as (1) and (2) are statistics with respect to the true values they also include any systematic errors.)
I mostly use \mar\ rather than RMS because the former is more robust.
If the residuals had a Gaussian distribution then the RMS would equal the Gaussian $1\sigma$ which is $\sqrt{\pi/2}=1.25$ times larger than \mar. But usually there are outliers which increase the RMS significantly beyond this.

\subsection{Uncertainty estimates}\label{sect:errest}

If vectors ${\bmath y}$ and ${\bmath x}$ are related by a transformation 
${\bmath y} = {\mathbfss A} {\bmath x}$
then a standard result of matrix algebra is that the covariance of 
${\bmath y}$ is ${\mathbfss C}_y = {\mathbfss A} {\mathbfss C}_x {\mathbfss A}^T$
where ${\mathbfss C}_x$ is the covariance of ${\bmath x}$. Applying this to equation~\ref{eqn:deltaap}
gives us an expression for the covariance in the APs
\begin{equation}
{\mathbfss C}_{\phi}  =  (\sens^T \sens)^{-1}\sens^T {\mathbfss C}_p  \sens (\sens^T \sens)^{-1}
\label{eqn:apcov}
\end{equation}
as a function of the sensitivity (calculated at the estimated APs) and the covariance in the measured photometry, ${\mathbfss C}_p$. 
(This equation assumes that \ilium\ provides unbiased AP estimates and that the
sensitivities have zero covariance. It can also be written ${\mathbfss C}_{\phi} = {\mathbfss M} {\mathbfss C}_p {\mathbfss M}^T$ where {\mathbfss M} is the update matrix introduced in equation~\ref{eqn:m}.)
${\mathbfss C}_p$ can be estimated from a photometric error model, and will be diagonal 
if the photometric errors in the bands are independent.
Even in this case ${\mathbfss C}_{\phi}$ is generally non-diagonal: the AP estimates are correlated on account of the sensitivities.  

Because we have a forward model we can calculate a goodness-of-fit (GoF)
for any estimate of the APs.
Here I simply use the reduced-\chisq\ to measure the difference between the observed spectrum and predicted spectrum\footnote{I use $I-1$ degrees of freedom rather than $I$ because all the spectra have a common G magnitude, so the bands are not all strictly independent.} 
\begin{equation}
{\rm GoF} = \frac{1}{I-1} \sum_{i=1}^{i=I} \left ( \frac{p_i - {\hat p_i}}{\sigma_{p_i}}  \right )^2
\label{eqn:gof_chisq}
\end{equation}
where ${\hat p_i} = f_i({\bmath \phi_j})$ is the forward model prediction and $\{ \sigma^2_{p_i} \} = {\rm diag}( {\mathbfss C}_p)$ is the expected photometric noise.  (Despite the name, a larger value refers to a poorer fit!) As the GoF can be measured without knowing the true APs, it can be used for detection of poor solutions or outliers.

Conventional methods of AP estimation via direct inverse modelling (e.g.\ with SVMs or ANNs) do not naturally provide uncertainty estimates and must usually resort to time-intensive sampling methods, such as resampling the measured spectrum according to its estimated covariance.
They cannot provide a GoF at all because they lack a forward model.

\subsection{Signal-to-noise weighted AP updates}

The update equation (\ref{eqn:deltaap}) only takes into account the sensitivity of the bands, not their SNR.  However, even if a band is very sensitive to an AP in principle, if its measurement is very noisy then it is less useful. We could accommodate this by including a factor proportional to ${\mathbfss C}_p^{-1}$ into equation~\ref{eqn:deltaap} which would down-weight noisier measurements. Preliminary results using this on the TG problem (see section~\ref{sect:datasets}) show it actually degrades performance at G=15, but gives some improvement at G=18.5 (Bailer-Jones~\citealp{cbj09c}).

\section{Assessing the algorithm}\label{data}

\subsection{Gaia simulations}\label{gaiasims}

To illustrate \ilium\ I apply it to estimate stellar APs from simulated Gaia stellar spectra and thereby also make preliminary predictions of the mission performance.
Gaia will observe all of its targets with two low-dispersion slitless prism spectrographs, together covering the wavelength range 
from 350--1050\,nm. (These are creatively called ``BP'' for blue photometer and ``RP'' for red photometer.) The dispersion varies from 3\,nm/pixel at the blue end to 30\,nm/pixel at the red end (Brown~\citealp{brown06}).  The blue and the red spectra are each sampled with 60 pixels, but as the line-spread-function of the spectrograph is much broader, these samples are not independent. After removing low SNR regions of the modelled spectra, I retain 34 pixels in BP covering 338--634\,nm and 34 pixels in RP covering 667--1035\,nm. This is a slightly narrower range (and 18 pixels fewer) than the one adopted by Bailer-Jones et al.~\citep{cbj08} for quasar classification with similar spectra.

\begin{figure}
\begin{center}
    \includegraphics[width=0.34\textwidth]{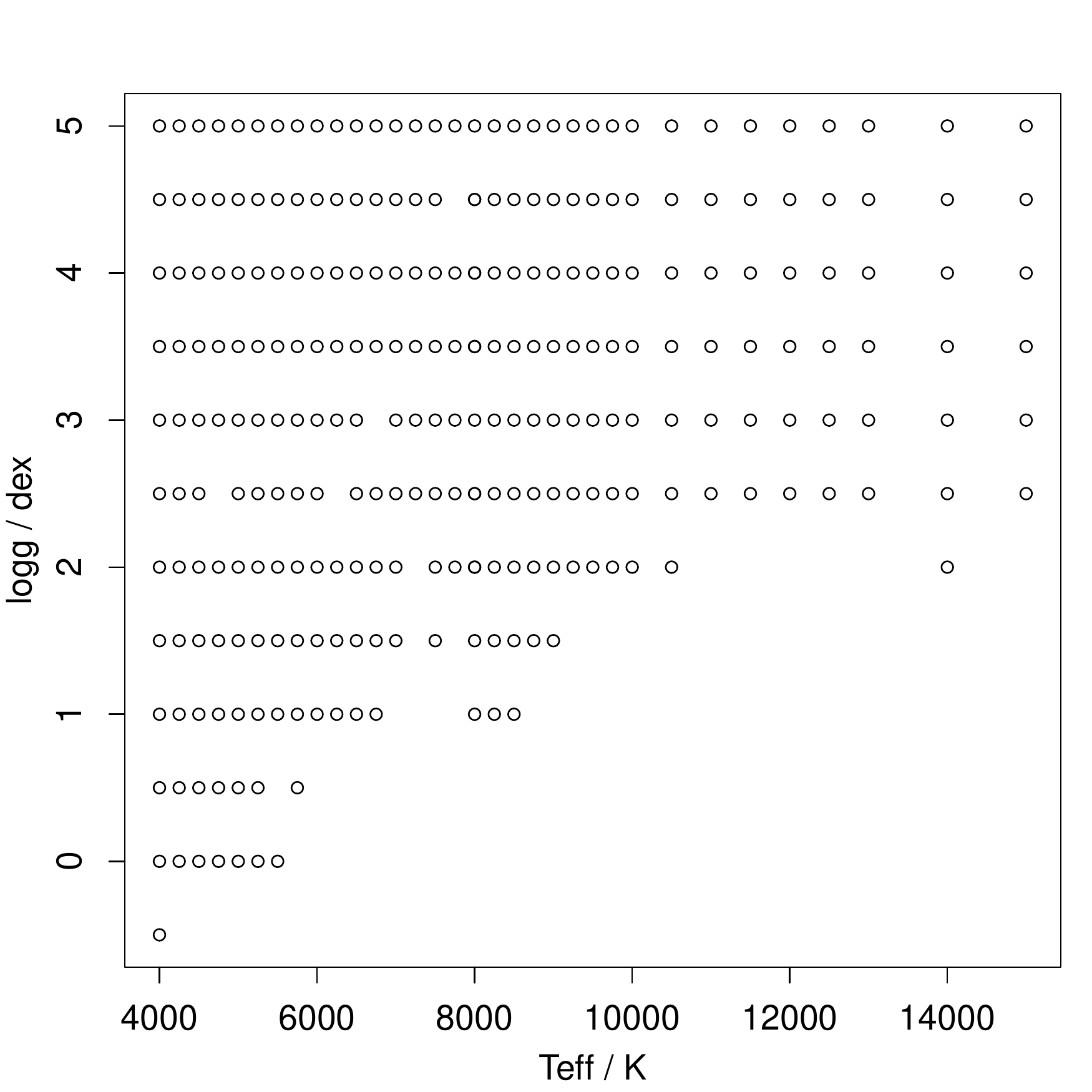} \caption{The \teff--\logg\ grid 
      of the data used in the experiments.}\label{fig:apgrid_teff_logg}
\end{center}
\end{figure}

Extensive libraries of BP/RP spectra have been simulated by the Gaia DPAC using the GOG (Gaia Object Generator; Luri et al.~\cite{luri05}, Isasi~\citealp{isasi09}) instrument model and libraries of input spectra.\footnote{For Gaia pundits: I use the CU8 cycle 3 simulations of the nominal (discrete) libraries (Sordo \& Vallenari~\citealp{sordo08})}
Here I use the Basel (Lejeune et al.~\citealp{lejeune97}) and Marcs (e.g.\ Gustafsson et al.~\citealp{gustafsson08}) stellar libraries.
The former (as used here) includes 17 \teff\ values from 8000--15\,000\,K with non-uniform spacing and the latter 17 \teff\ values from 4000--8000\,K in uniform steps of 250\,K. (Together there are 33 unique \teff\ values because 8000\,K is in both.) Together the grids span \logg\ values from $-0.5$ to $5.0$\,dex in steps of $0.5$\,dex, although the grid is incomplete for astrophysical reasons (Fig.~\ref{fig:apgrid_teff_logg}).  \feh\ ranges from $-4.0$\,dex to $+1.0$\,dex with 13 discrete values for the cooler stars (with \teff\,$\leq 8000$\,K): The spacing is 0.25\,dex from $+1$ to $-1$, followed by points at $-1.5$, $-2.0$, $-3.0$ and $-4.0$. Not all \feh\ are present at all \teff--\logg\ combinations. Each star has been simulated at one of 
ten values of the interstellar extinction, \av\,$\in \{0, 0.1, 0.5, 1, 2, 3, 4, 5, 8,10\}$ with \rv\,=\,3.1.  (Note that \av\ is the extinction parameter defined by Cardelli et al.~\citealp{cardelli89}. It is not the extinction in the $V$ band.)
The Marcs library additionally shows variation in the alpha element abundances, \aabun, representing five values from 0.0--0.4\,dex in 0.1\,dex steps. Hence this combined library shows variance in five APs, \teff, \av, \logg\, \feh, \aabun, the first two of which are ``strong'' and the latter three ``weak''.  The total number of spectra is 46\,310. \ilium\ will be used to estimate the first four APs; \aabun\ will be ignored and contributes cosmic scatter.

\begin{figure}
\begin{center}
\includegraphics[width=0.47\textwidth]{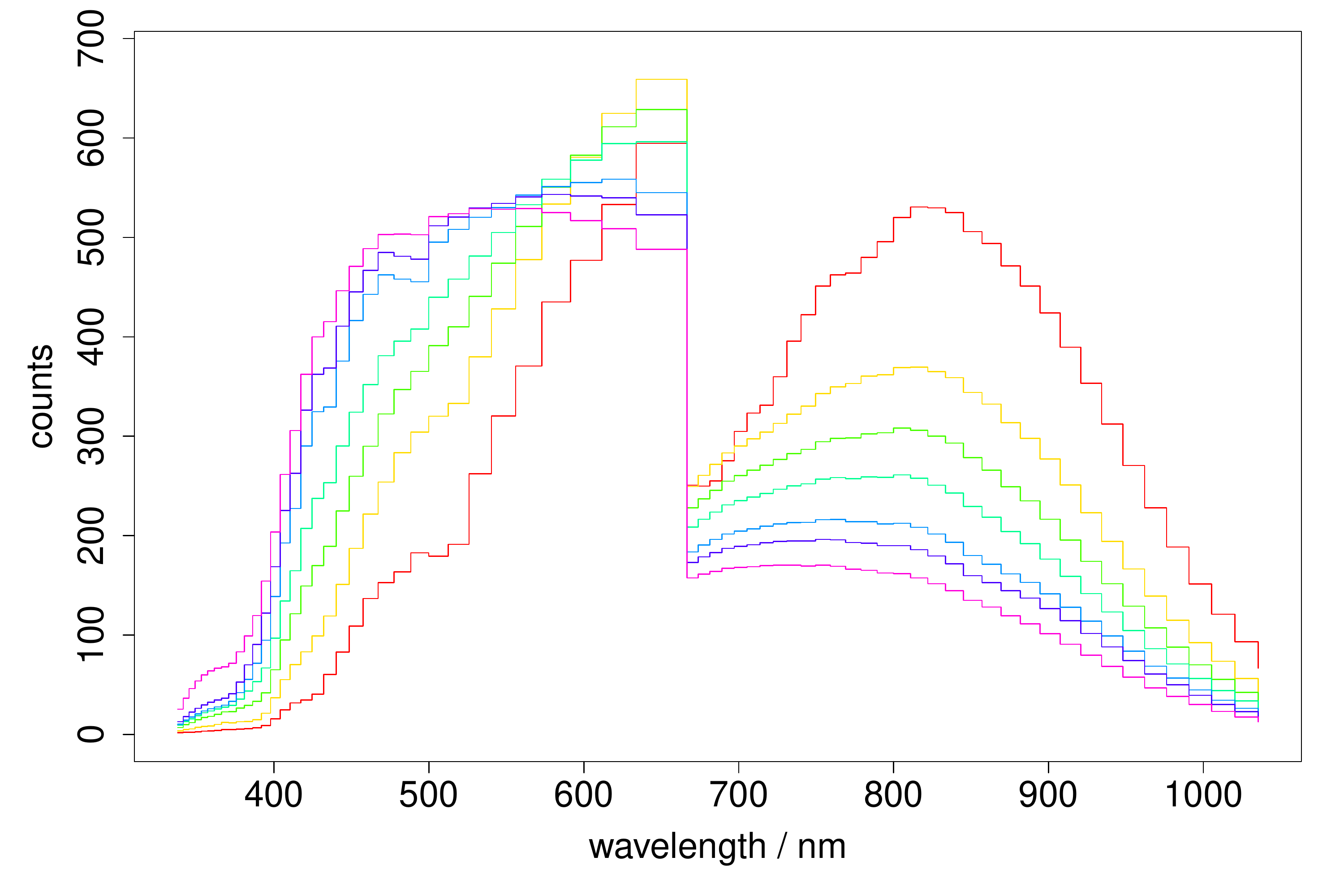}
    \caption[]{Noise-free Gaia spectra for solar metallicity dwarfs at zero extinction with \teff\,=\,\{4000, 5000, 6000, 7000, 8500, 10000, 15000\}\,K, increasing monotonically from bottom (red) to top (violet) at long wavelengths. 
They are composed of two spectra (BP and RP) to the left and right of about 660\,nm.
The ordinate is in units of photoelectrons (to within some constant multiple), not energy flux. 
\label{fig:teff_spectra}}
\end{center}
\end{figure}

\begin{figure}
\begin{center}
\includegraphics[width=0.47\textwidth]{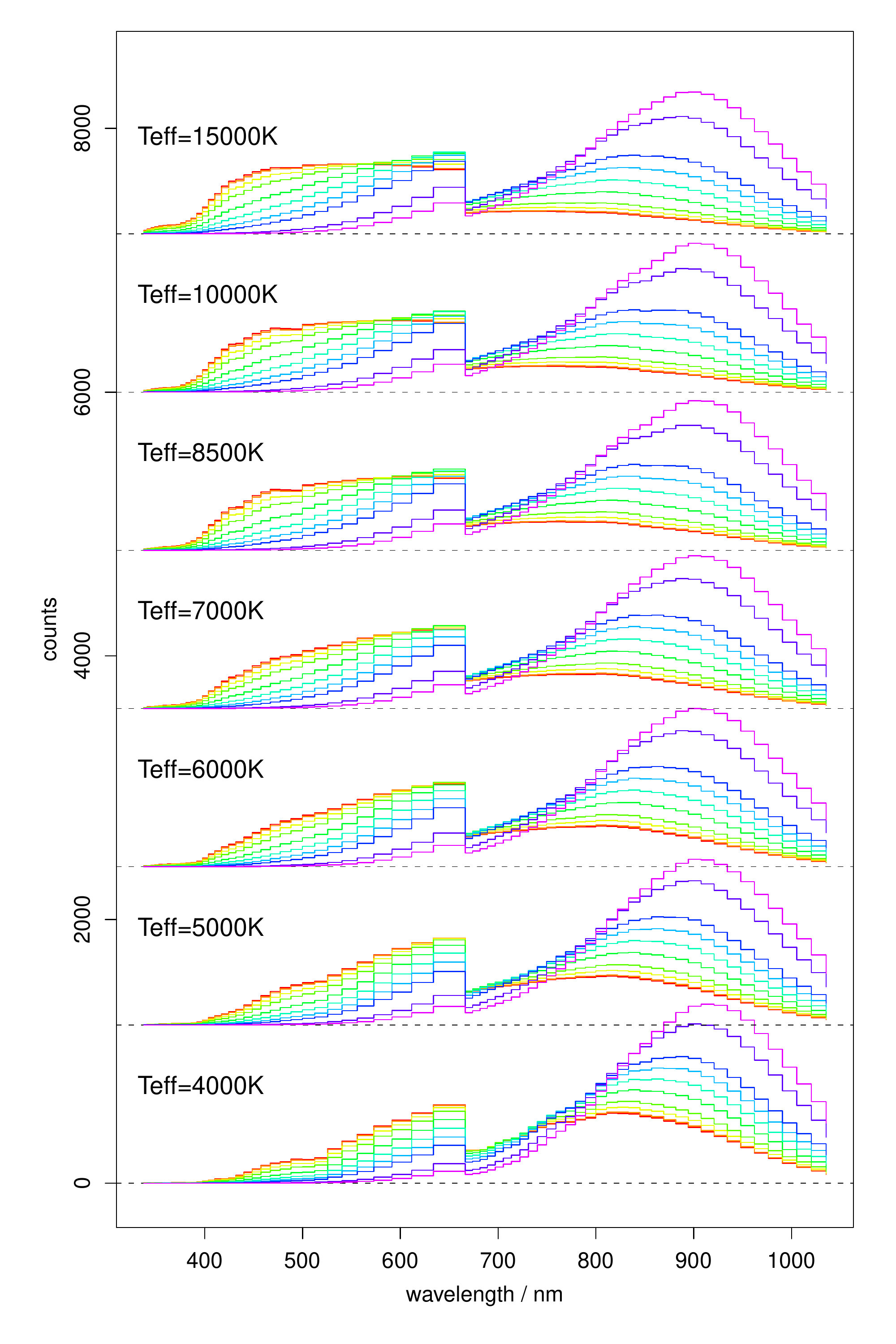}
    \caption[]{Noise-free Gaia spectra at a range of \teff\ (as in Fig.~\ref{fig:teff_spectra}) and 
\av\
 (0.0, 0.1, 0.5, 1, 2, 3, 4, 5, 8, 10)
ranging from 0.0\,mag (lowest line at long wavelengths; in red) to 10.0\,mag (highest line at long wavelengths; in violet). Each temperature block has been offset by 1200 counts for clarity (the zero levels are shown by the dashed lines).
\logg\,=\,4.0\,dex, \feh=\,0.0\,dex and \aabun\,=\,0.0\,dex in all cases.
\label{fig:teff_av_spectra}}
\end{center}
\end{figure}

\begin{figure}
\begin{center}
    \includegraphics[width=0.47\textwidth]{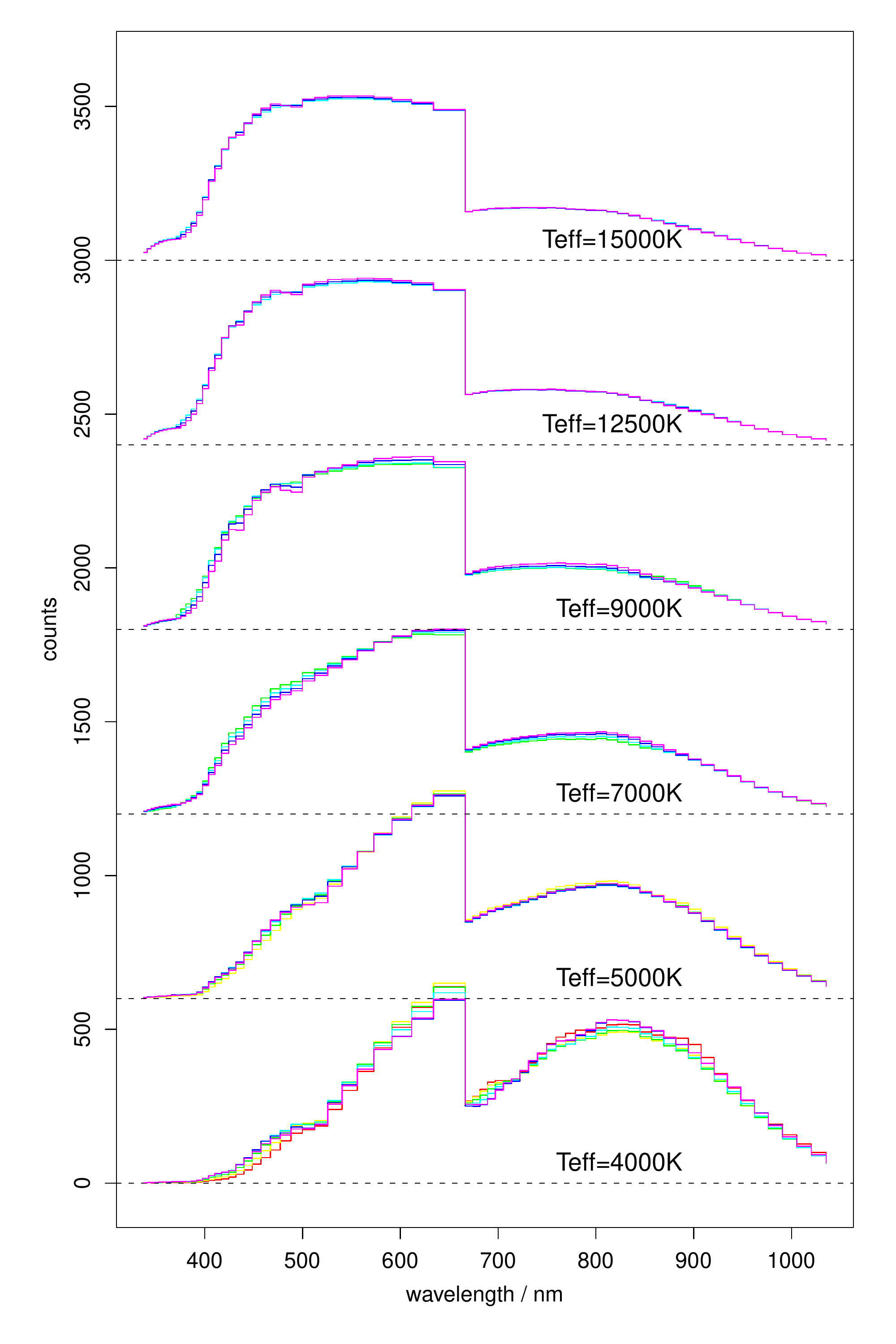}
    \caption[]{Noise-free Gaia spectra at a range of \teff\ and 
\logg\ $(-0.5,0.5,2,3,4,5)$, with the lowest gravity (red) forming the lowest curve at the red end of the spectrum and the highest gravity (violet) the highest. Note that not all gravities are present at all \teff\ due to the limitations of reality.
Each temperature block has been offset by 600 counts for clarity (the zero levels are shown by the dashed lines)
\av\,=\,0.0\,mag, \feh=\,0.0\,dex and \aabun\,=\,0.0\,dex in all cases.
\label{fig:teff_logg_spectra}}
\end{center}
\end{figure}

\begin{figure}
\begin{center}
    \includegraphics[width=0.47\textwidth]{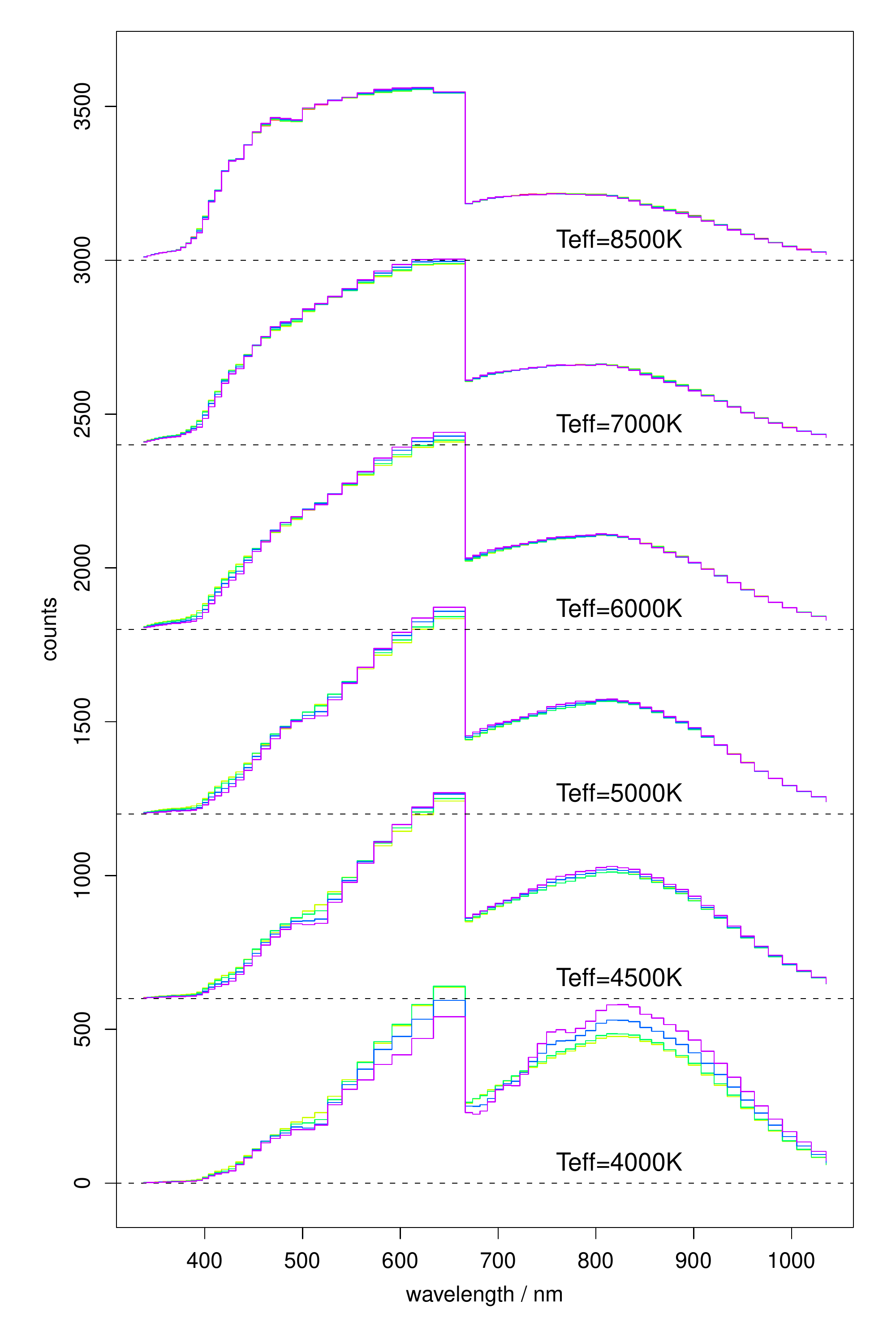}
    \caption[]{Noise-free Gaia spectra at a range of \teff\ and 
\feh\ $(-3,-2,-1,0,+0.5)$, with the lowest metallicity (red) forming the lowest curve at the red end of the spectrum and the highest metallicity (violet) the highest. Note that not all metallicities are present at all \teff\ due to limitations of the simulated libraries.
Each temperature block has been offset by 600 counts for clarity (the zero levels are shown by the dashed lines)
\av\,=\,0.0\,mag, \logg\,=\,4.0\,dex and \aabun\,=\,0.0\,dex in all cases.
\label{fig:teff_feh_spectra}}
\end{center}
\end{figure}

GOG simulates the number of photoelectrons (``counts'') in the spectral bands. 
(While these will be calibrated in physical flux units before being published to the community, the classification work by DPAC is currently done in photoelectron space.)
The variance in the spectra due to the four APs of interest is demonstrated in the example spectra plotted in Figs.~\ref{fig:teff_spectra}, \ref{fig:teff_av_spectra}, \ref{fig:teff_logg_spectra} and \ref{fig:teff_feh_spectra}. The first plot shows the variance due to \teff\ only. 
The break between the BP and RP instruments around 660\,nm is clear, as is the highly variable dispersion. The lower counts in RP immediately to the right of the break compared to BP is primarily due to the higher dispersion (fewer photons per band).
In the other plots two APs are varied while the other two are held constant. Fig.~\ref{fig:teff_av_spectra} demonstrates the strong impact of both \teff\ and \av\ variations. (As I will discuss in section~\ref{degeneracy}, these two APs are highly degenerate in these data.)  The third and fourth figures clearly demonstrate why \logg\ and \feh\ are weak parameters: they have little impact on the spectra compared to the variance due to \teff\ or \av. In particular, at high temperatures the spectra show essentially no sensitivity to \feh.  

\begin{figure}
\begin{center}
    \includegraphics[width=0.35\textwidth]{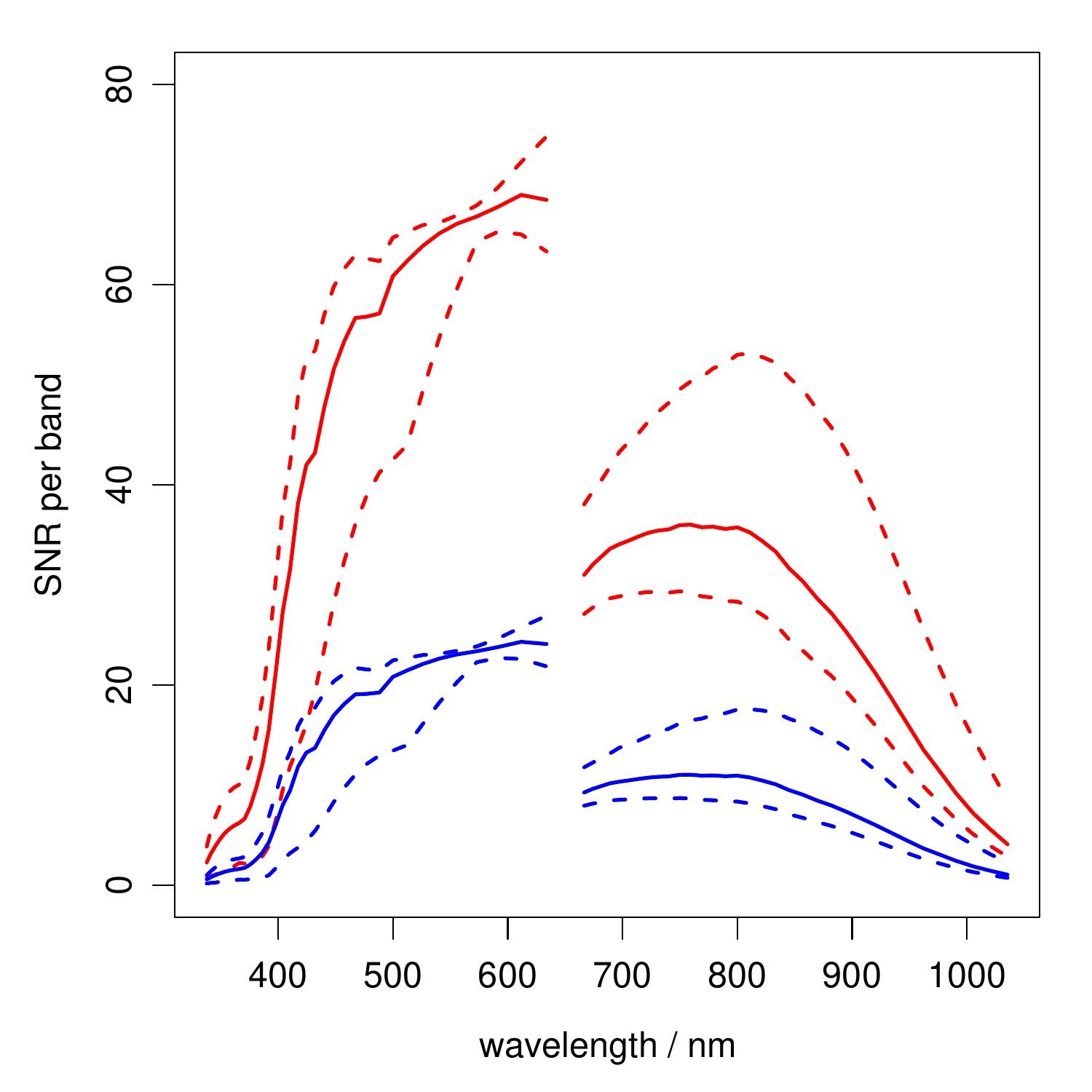} 
    \caption{The median signal-to-noise ratio (SNR) (solid line) and 0.1 and 0.9 quartiles
      (dashed lines) across the set of zero extinction dwarf stars for G=18.5 (red/thicker lines) and
      G=20.0 (blue/thinner lines). Compared to the G=18.5 curve, the SNR at G=15 is 8--6 times larger between 400 and 660\,nm and 7--12 times larger between 660 and 1000\,nm.}
\label{fig:snr_dwarfs}
\end{center}
\end{figure}

The AP estimation accuracy is, of course, a strong function of the SNR in the spectrum.  As Gaia has a fixed sky scanning law, the SNR depends on the source magnitude and the number of observations (because the individual observations are combined into a single end-of-mission spectrum). I adopt here 72 observations for all spectra (the mean number of observations per source for a 5-year mission), and report results as a function of just the magnitude.  The simulator noise model takes into account the source, background and background-subtraction photon (Poisson) noise as well as the CCD readout noise.  At present errors due to the combination of spectra, charge-transfer inefficiency and CCD radiation damage are not explicitly included. There is instead a factor to account for general processing and calibration errors.  All of these noise terms are combined into a zero mean Gaussian model for each pixel, the standard deviation of which is a function of the G-band apparent magnitude.  (The G-band is the filterless band -- defined by the mirror and CCD response -- in which the Gaia astrometry is obtained, covering a range similar to BP/RP.) This defines a ``sigma spectrum'' for each star from which I generate random numbers in order to simulate noisy spectra at G=15, 18.5 and 20. The resulting SNR is a strong function of wavelength and the specific source, and is summarized in Fig.~\ref{fig:snr_dwarfs}.

\subsection{Data sets and forward model fitting}\label{sect:datasets}

In the following sections I will apply Gaia to four distinct problems according to the APs we are trying to determine. In each case the forward model is fit to a grid varying in at least those APs, and in some cases the grid also shows variance in another AP (which therefore acts to provide additional ``cosmic scatter'').  For each case the \ilium\ forward model is fit (trained) using the full, noise-free data set over the complete range of the represented APs.
The corresponding noisy data set (at G=15, 18.5 or 20) is split randomly into two equal halves: one half is used for initialization (selecting the nearest neighbour) and \ilium\ is applied to the other half (or a random subset of it where indicated below) on which the performance is evaluated.\footnote{These target spectra must first be normalized to have the same
counts level as those on which \ilium\ was trained. For this I just use the G magnitude to scale the counts. However, as the G-band is not identical to the BP/RP band adopted, this gives rise to a normalization offset 
between spectra even for a common G magnitude. For example, over the TG grid the integrated BP/RP counts varies by up to 10\%, dependent primarily on \teff. A better normalization might be ``area normalization'', i.e.\ dividing the counts in each band by the sum over all bands for that spectrum. My G-band normalization is in principle conservative, as it mimics a small calibration bias.}
(The legitimacy of this procedure for evaluating performance is discussed in appendix~\ref{sect:modelassess}.)
In addition to reporting global results (over the full AP ranges present in the training set) I also measure performance on subsets of the evaluation data set, in particular the \feh\ just for cool stars (the \ilium\ forward model is {\em not} refit).  The problems and their grids are as follows (the name indicates the APs being determined: Temperature, Gravity, Metallicity and/or Extinction)
\begin{itemize}
\item TG: Estimation of \teff$+$\logg\ (1 strong and 1 weak AP), for stars with \av\,=\,0 and \feh\,=\,0, some 274 stars. I also build a second TG model (TG-allmet) which is trained and evaluated with the full \feh\ range in the grid ($-4$ to $+1$\,dex). This contains 4361 stars, of which a quarter are used in the evaluation set.
\item TM: Estimation of \teff$+$\feh\ (1 strong and 1 weak AP), for stars with \av\,=\,0 and either \logg\,$\in\!\{4.0, 4.5, 5.0\}$ (TM-dwarfs; 1716 stars), or \logg\ $\in\!\{1.0, 1.5, 2.0, 2.5, 3.0\}$, (TM-giants; 1882 stars). I also build a third model (TM-allgrav) which is trained and evaluated with the full \logg\ range in the grid ($-0.5$ to $+5$\,dex). This contains 4361 stars (it's the same grid as used for TG-allmet of course), of which a quarter are used in the evaluation set.
\item TAG: Estimation of (\teff, \av)$+$\logg\ (2 strong and 1 weak AP), for stars with \feh\,=\,0. This has 2740 stars, of which a random selection of 1000 is used for evaluation.
\item TAM: Estimation of (\teff, \av)$+$\feh\ (2 strong and 1 weak AP), for dwarfs with \logg\,$\in\!\{4.0, 4.5, 5.0\}$. This has 17\,160 stars of which a random selection of 1000 is used for evaluation.
\item TGM: Estimation of \teff$+$(\logg, \feh) (1 strong and 2 weak APs, for stars with \av\,=\,0.
 This has 4361 stars of which a random selection of 1000 is used for evaluation.
\end{itemize}
In appendix~\ref{sect:compare} the \ilium\ results are compared with results from an SVM on some of these problems.


\begin{table*}
\begin{center}
\caption{Summary statistics (defined in section~\ref{errstat}) of the \ilium\ performance on various experiments.
The first column indicates the model and grid used (defined in section~\ref{sect:datasets}). The second column defines the magnitude and composition of the evaluation data set: F means the full AP ranges (as present in the training set); L means only stars with \teff\,$\leq$\,7000\,K.
``$<$'' in the \mr\ column indicates that the systematic error has a magnitude less than twice the standard error in the mean (\absmr$\,\leq 2\!\times\!1.25$\,\mar/$\sqrt{N}$), i.e.\ is statistically insignificant. (Quite a few not so marked are only marginally significant.)
The units of the error statistics are logarithmic for all variables (dex for all except \av\ which is in magnitudes) and are on the proper variables (not the standardized variables).
Multiplying by 2.3 converts to fractional errors for \logteff. }\label{sumres}
\begin{tabular}{ll*{12}{r}}
\hline
model & evaluation & \multicolumn{3}{c}{\av}        & \multicolumn{3}{c}{\logteff}        & \multicolumn{3}{c}{\logg}        & \multicolumn{3}{c}{\feh} \\ \cline{3-5} \cline{9-11}
           & sample      & \mr   & \mar   & \rms          & \mr   & \mar   & \rms                 & \mr   & \mar   & \rms             & \mr   & \mar   & \rms \\
\hline
TG & F G=15      & & &                               & $<$ & 0.0010 & 0.0014                         & $<$ & 0.065 & 0.093    & & & \\
TG & F G=18.5   & & &                               & $<$ & 0.0057 & 0.0080                         & $<$ & 0.35 & 0.51          & & & \\
TG & F G=20      & & &                               & 0.0045 & 0.019 & 0.027                        & $<$ & 1.14 & 1.53            & & & \\
TG-allmet & F G=15  & & &                      & $7.4e^{-4}$ & 0.0058 & 0.0078               & $<$ & 0.53 & 0.90    & & & \\ 
TG-allmet & F G=18.5  & & &                   &  $8.2e^{-4}$ &  0.0083 & 0.011               &  $<$ & 0.66 & 0.99     & & & \\ 
TG-allmet & F G=20  & & &                      & 0.0025 & 0.019 & 0.027                         & $-$0.14 & 1.19 & 1.59    & & & \\ 
\hline
TM-dwarfs & L G=15    & & &                   & $<$ & 0.0017 & 0.0020     & & &          & $<$ & 0.14 & 0.24 \\ 
TM-dwarfs & L G=18.5 & & &                  & $<$ & 0.0024 & 0.0033      & & &          & $-$0.037 & 0.26 & 0.42  \\
TM-dwarfs & L G=20   & & &                   & $<$ & 0.0070 & 0.0090      & & &          & $<$ & 0.82 & 1.14  \\
TM-giants & L G=15    & & &                   & $<$ & 0.0028 & 0.0037     & & &           & $<$ & 0.22 & 0.34 \\
TM-giants & L G=18.5 & & &                   & $3.4e^{-4}$ & 0.0035 & 0.0045     & & &     & $<$ & 0.31 & 0.50  \\
TM-giants & L G=20   & & &                    & $0.0011$ & 0.0073 & 0.0092       & & &      & $-$0.11 & 0.74 & 1.08  \\
TM-allgrav & L G=18.5   & & &                & $<$ & 0.0052 & 0.0070          & & &           & $<$ & 0.40 & 0.60  \\
TM-allgrav & F G=18.5   & & &                & 0.0031 & 0.012 & 0.017         & & &           & & & \\
\hline
TAG & F G=15         & $-$0.0067 & 0.072 & 0.15    & 0.0014 & 0.013 & 0.026    & $-$0.052 & 0.29 & 0.57   & & & \\
TAG & F G=18.5      & 0.039 & 0.30 & 0.45              & 0.017 & 0.061 & 0.094      & $-$0.22 & 1.10 & 1.53     & & & \\
TAM & L G=15      & 0.032 & 0.18 & 0.32                & 0.0022 & 0.018 & 0.029    & & &    & $<$ & 0.46 & 0.79 \\
TAM & L G=18.5   & 0.047 & 0.52 & 0.74                & 0.010 & 0.056 & 0.080      & & &    & $-$0.38 & 1.34 & 1.81 \\
\hline
TGM & L G=15     & & &   & $8.3e^{-5}$ & 0.0007 & 0.0012   & $-$0.022 & 0.15 & 0.26      & 0.014 & 0.084 & 0.19 \\
TGM & L G=18.5  & & &   & $8.8e^{-4}$ & 0.0043 & 0.0060   & $-$0.11   & 0.84 & 1.15      & $-$0.045 & 0.35 & 0.60 \\
TGM & F G=18.5  & & &   & $2.3e^{-3}$ & 0.0089 & 0.014    & $-$0.06   & 0.59 & 0.86      &  & &  \\
\hline
\end{tabular}
\end{center}
\end{table*}


\section{Application to the \teff+\logg\ problem (TG)}\label{sect:tefflogg}

\begin{figure*}
\begin{center}
\includegraphics[width=0.70\textwidth, angle=0]{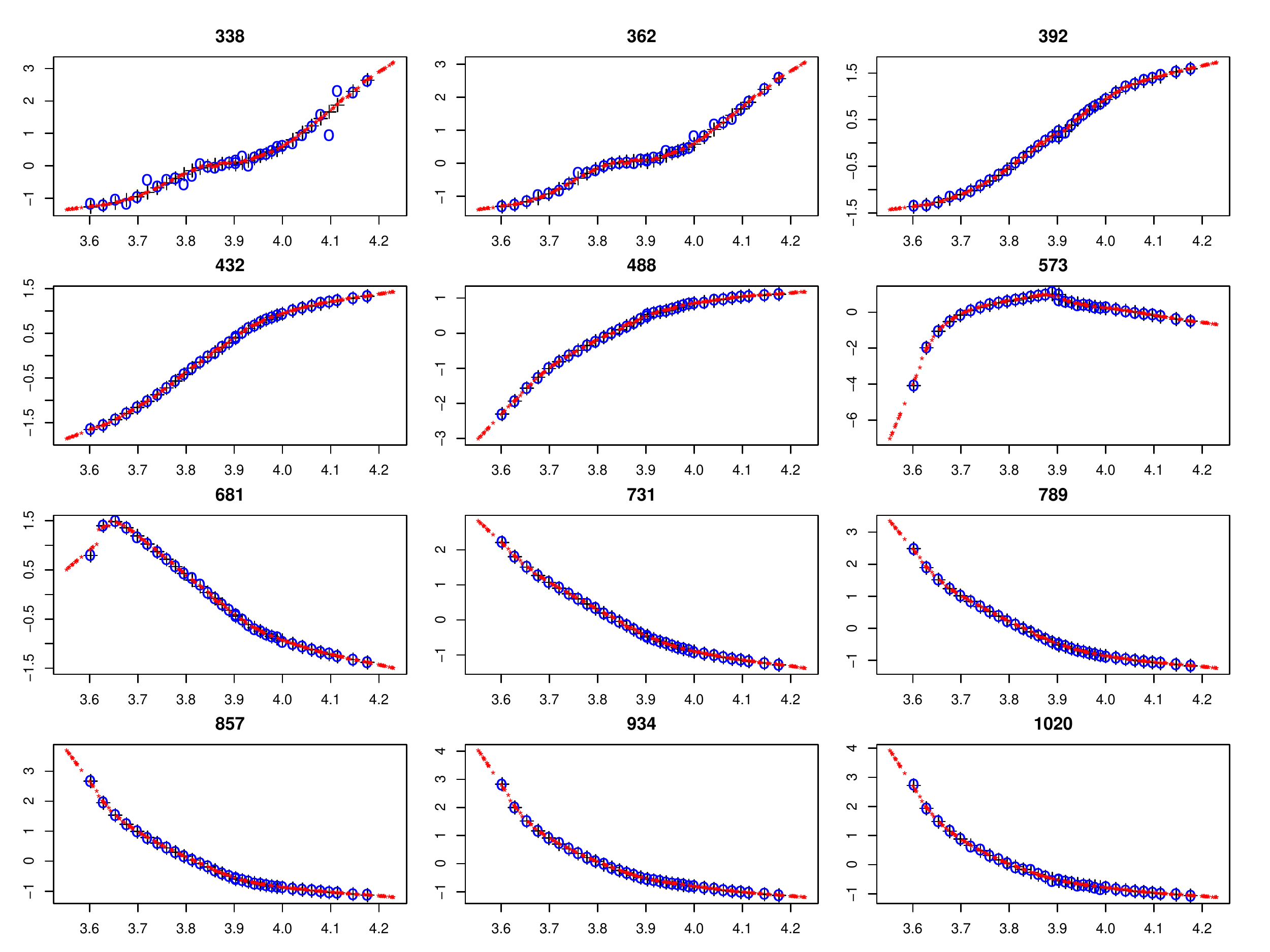}
\caption{Predictions of the full forward model for the TG problem as a function of
  \logteff\ with \logg\ fixed at 4.0\,dex for 12 different bands (the central wavelength of which
  in indicated at the top of each panel in nm). The black crosses are
  the (noise-free) grid points, the small red stars are the forward model
  predictions (at randomly selected AP values) and the blue circles
  the noisy (G=15) grid points. The photon counts plotted on the ordinate are in 
  standardized units.} \label{fig:formod_logg=4}
\end{center}
\end{figure*}

\begin{figure*}
\begin{center}
\includegraphics[width=0.70\textwidth, angle=0]{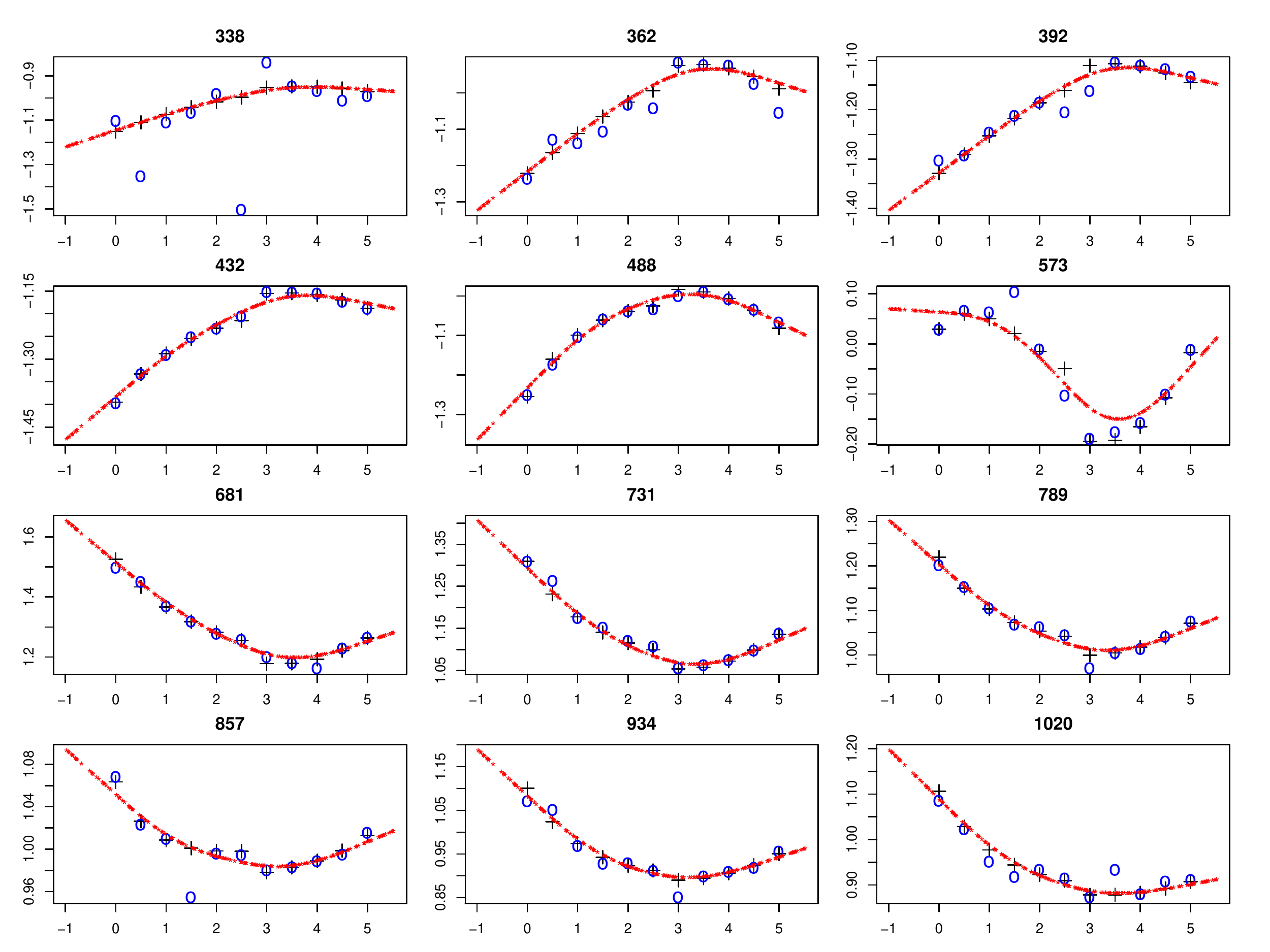}
\caption{As Fig.~\ref{fig:formod_logg=4}, but now showing predictions
  of the full forward model as a function of \logg\ at constant
  \teff=5000\,K.} \label{fig:formod_teff=5000}
\end{center}
\end{figure*}

First the forward model is fit for each band to the \teff--\logg\ grid shown in Fig.~\ref{fig:apgrid_teff_logg}. 
As a reminder, each forward model comprises the 1D function over \teff\ (the strong component) and 33 1D functions in \logg\ (the weak components), as described in section~\ref{sect:forward_model}. All of these are smoothing splines (section~\ref{fmfunctions}).  Figs.~\ref{fig:formod_logg=4} and~\ref{fig:formod_teff=5000} show the forward model fit for 12 bands at cuts of constant \logg\ and \teff\ respectively.  The fits are good: they show the degree of smoothness we would expect for these data plus a robust extrapolation. If we compare the flux scales between Figs.~\ref{fig:formod_logg=4} and~\ref{fig:formod_teff=5000} (these are standardized variables) we see how small the flux variation is as \logg\ varies over its full range compared to \teff: this is what it means to be a weak AP. Note also the small discontinuity in some of the bands in Fig.~\ref{fig:formod_logg=4} at 8000\,K (\logteff\,=\,3.903) where the Marcs and Basel libraries join.

\begin{figure}
\vspace*{-3ex}
\begin{center}
\includegraphics[width=0.45\textwidth, angle=0]{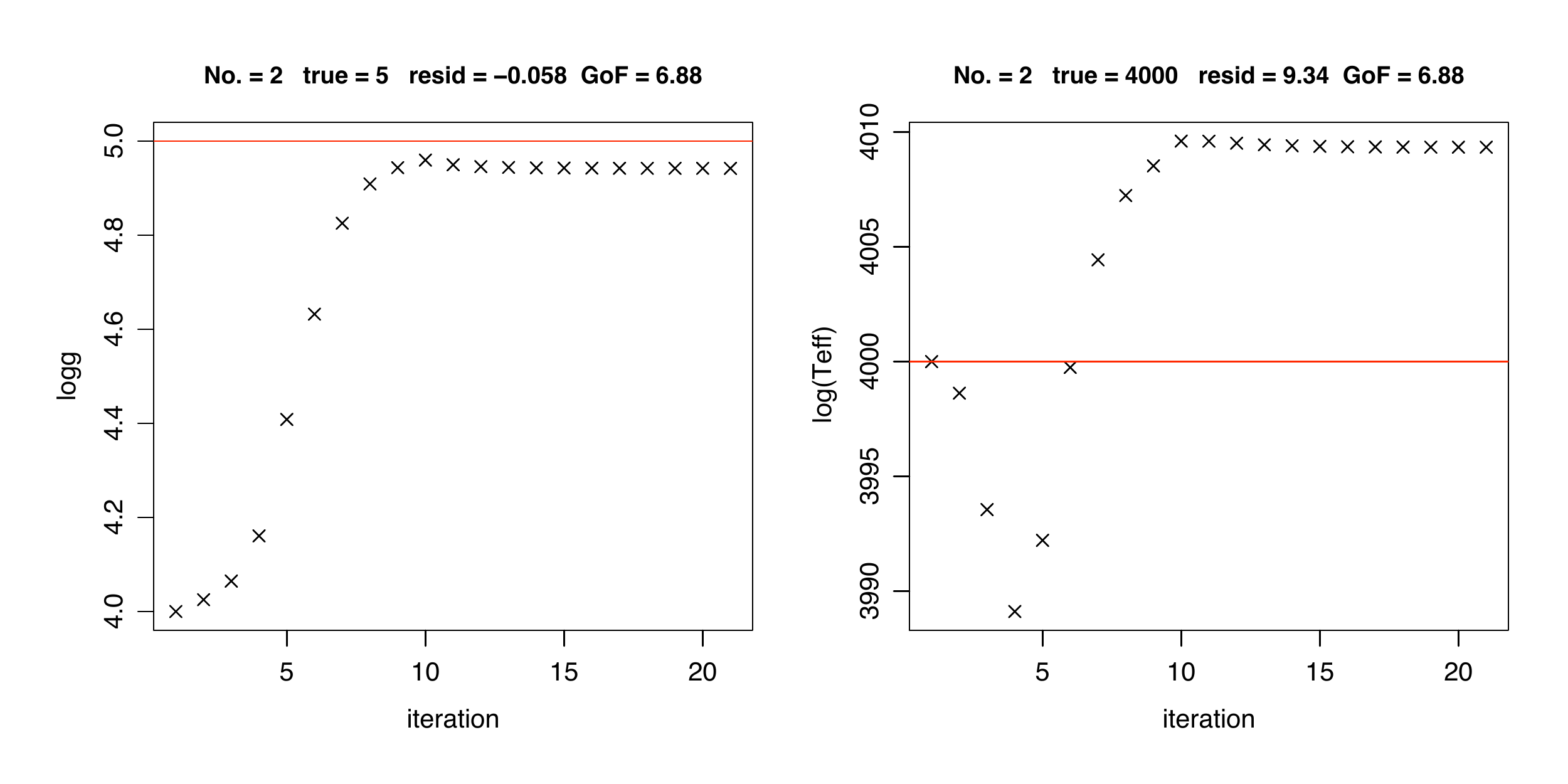}\vspace*{-3ex}
\includegraphics[width=0.45\textwidth, angle=0]{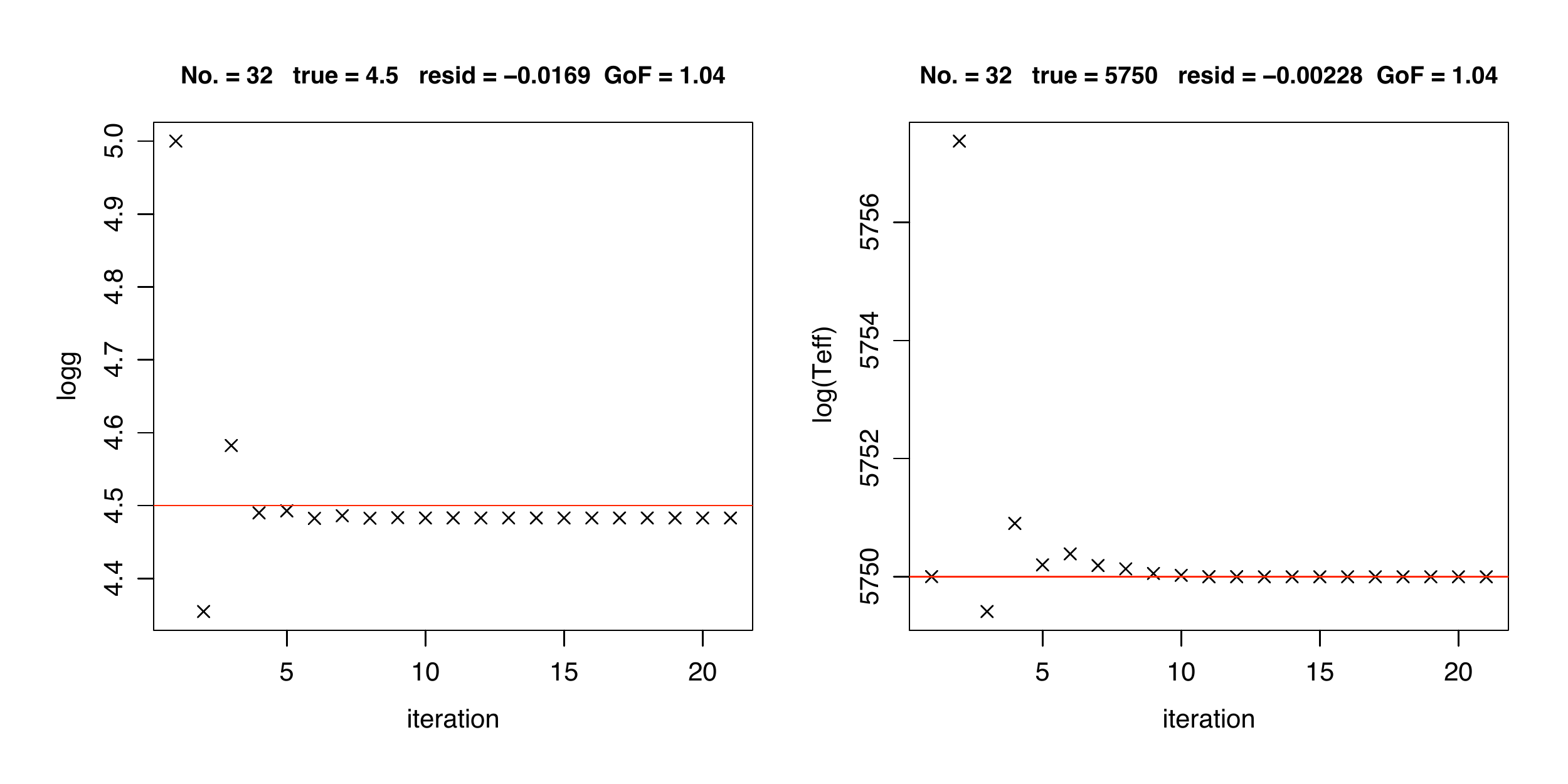}\vspace*{-3ex}
\includegraphics[width=0.45\textwidth, angle=0]{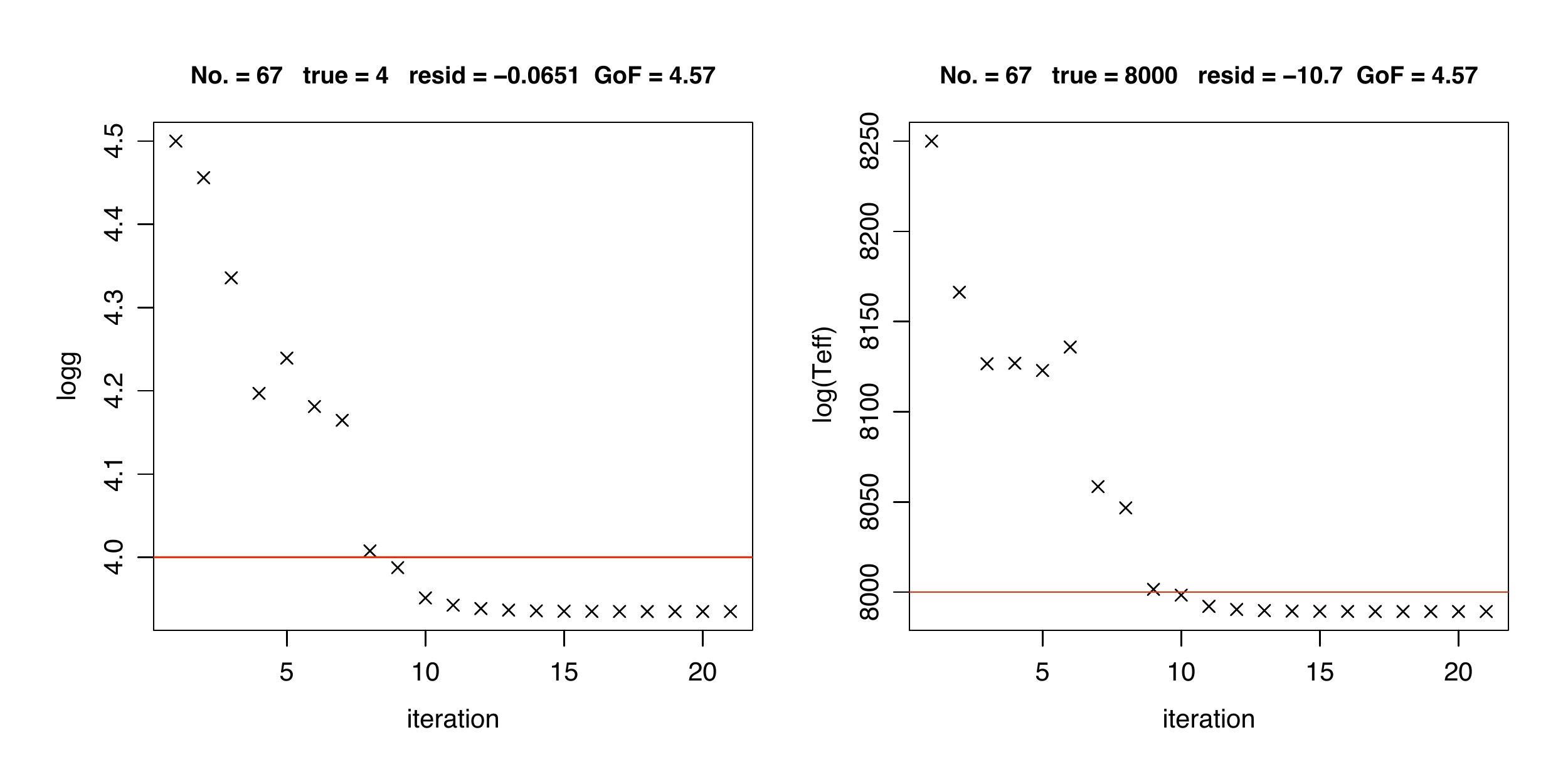}\vspace*{-3ex}
\includegraphics[width=0.45\textwidth, angle=0]{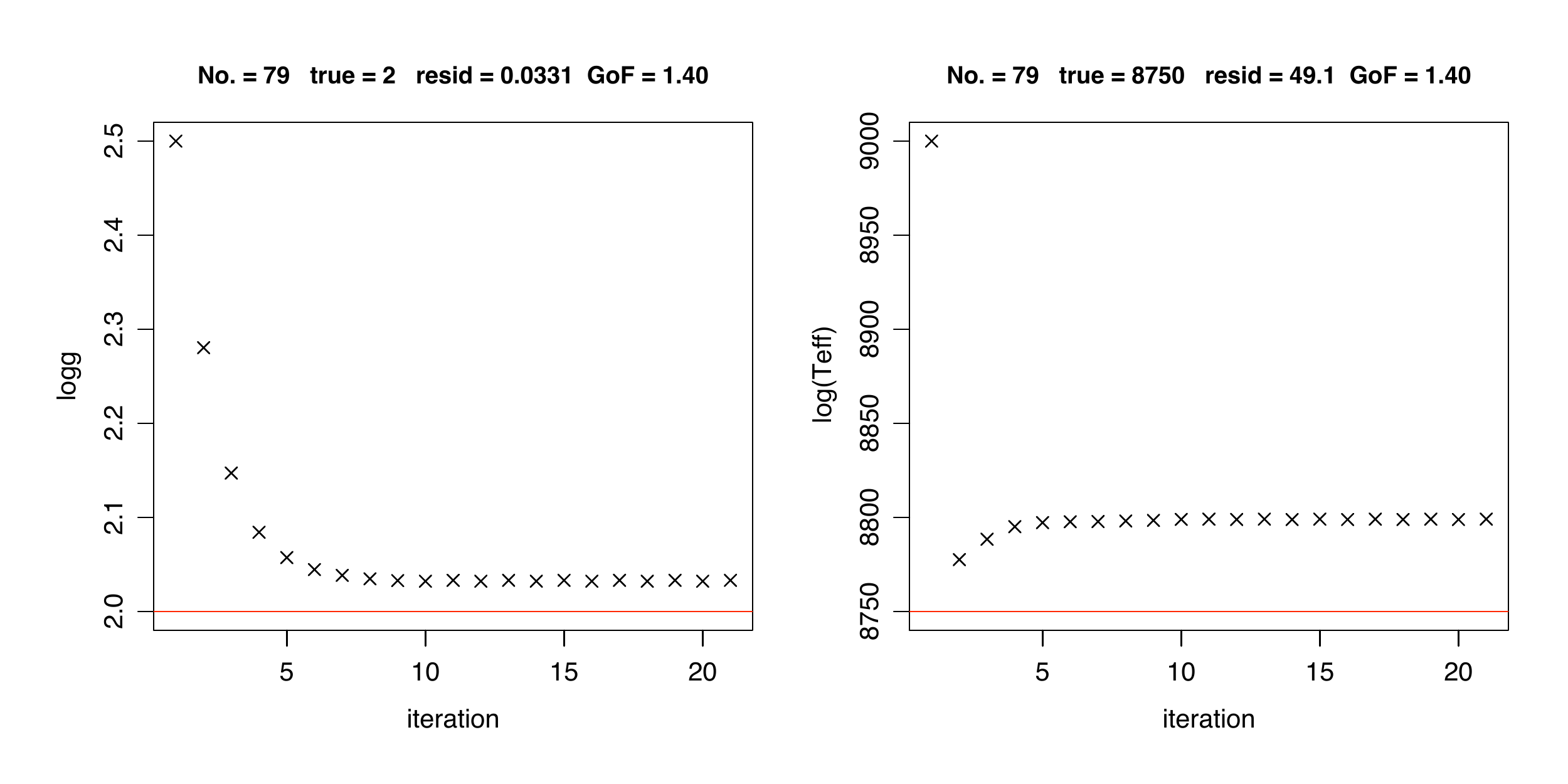}\vspace*{-3ex}
\includegraphics[width=0.45\textwidth, angle=0]{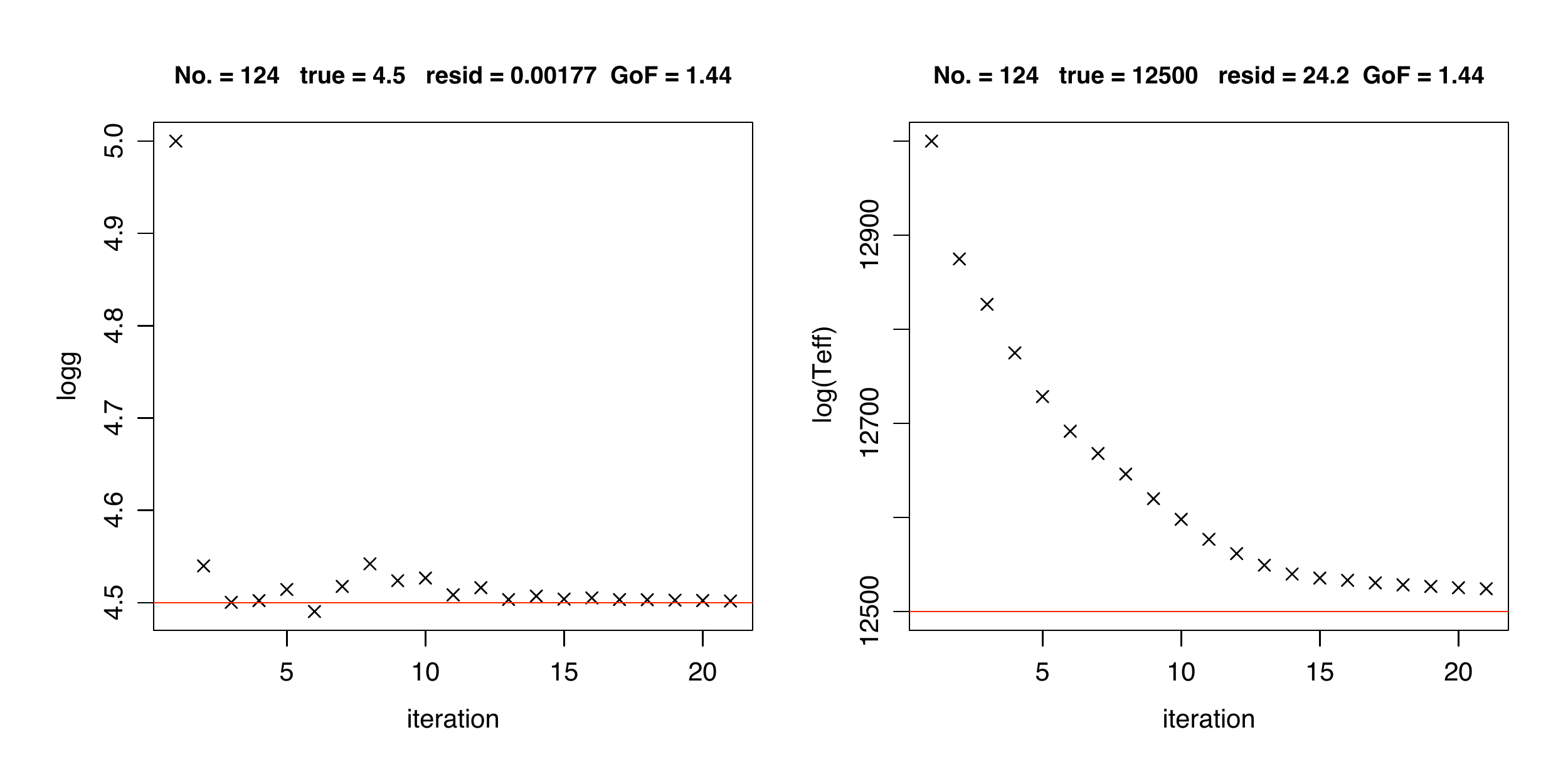}
\caption{AP evolution for 5 stars in the evaluation data set at G=15 (\logg\
  left, \teff\ right) for the TG problem. The true APs are written at the top of
  each panel pair and plotted as the red horizontal line. GoF is the
  reduced \chisq\ goodness-of-fit
  (equation~\ref{eqn:gof_chisq}).} \label{fig:test10.iter}
\end{center}
\end{figure}

Having trained \ilium\ I apply it to the evaluation data set at G=15.  Fig.~\ref{fig:test10.iter} shows five examples of how the AP estimates evolve. The first iteration is the nearest neighbour initialization; the final is the adopted AP estimate. The red (horizontal) lines show the true APs. Looking at many examples we see a range of convergence behaviours. Sometimes convergence is rapid, for other stars it takes longer. Sometimes it is smooth, other times not. It can be quite different for the two APs for a given star and depends also on the specific spectrum (which is noisy). Sometimes the nearest neighbour estimate is the correct one, and \ilium\ may actually iterate away from this and converge on a different value. Convergence (on something) is almost always achieved on this problem, even though there is nothing adaptive in the algorithm. This is an encouraging property.  Limit cycles are also seen in a handful of cases, but with negligible amplitudes on this problem (high SNR). For this specific problem, the extrapolation limits on the AP estimations (section~\ref{sect:apextrapolate}) and the limits on the AP updates at each step (section~\ref{sect:apupdates}) never had to be enforced by the algorithm.

\begin{figure}
\begin{center}
\hspace*{-0.5cm}
\includegraphics[width=0.54\textwidth]{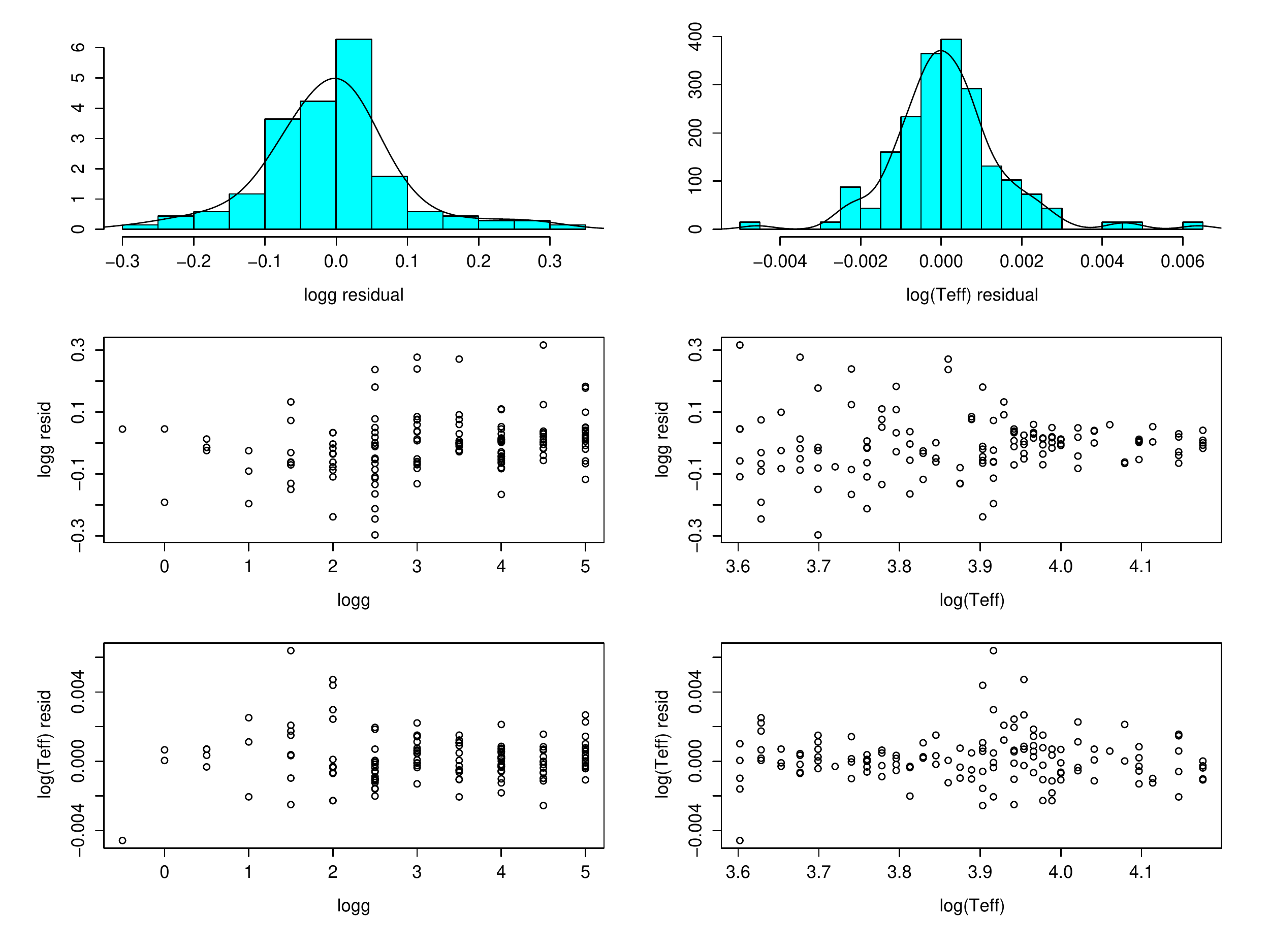}
\caption{AP residuals for the TG problem at G=15} \label{fig:itup3_test10_itup_residuals}
\end{center}
\end{figure}

Fig.~\ref{fig:itup3_test10_itup_residuals} plots the \ilium\ residuals (estimated minus true APs) on the 137 stars in the evaluation set; the statistics are summarized in line 1 of Table~\ref{sumres}. We see that the APs can be estimated very accurately (no significant systematic error, \mr) and very precisely (low scatter, \mar\ or \rms). Applying the same model to noisier spectra obviously increases the errors (lines 2 and 3 in the table), but at G=18.5 the mean absolute errors are still an acceptable 1\% in \teff\ and 0.35\,dex in logg. At G=20 \logg\ cannot be estimated accurately enough to reliably distinguish dwarfs and giants, although \teff\ is still okay with an expected error of 260\,K at 6000\,K, for example.  

This evaluation set is relatively small so the error statistics are subject to variation. Applying \ilium\ to 
an ensemble of randomly selected evaluation sets (at G=18.5), we see that the error statistics vary by 5--10\% (inter-quartile range) of their mean. The reported values are therefore reasonably representative.

The performance of an SVM on these TG problems is given in Table~\ref{svmres}.  For both APs and all three magnitudes \ilium\ is similar to or significantly better than SVM. (As there is some variance in the results from both methods I only consider the performance significantly different if the better one has a \mar\ at least 25\% smaller.) So in this limited variance problem, at least, the forward modelling approach improves performance.

It is hard to fairly compare the performance with nearest neighbours (NN), because on the one hand NN is limited by the density of its template grid, but on the other hand it can report the {\em exact} AP values. If we used a full, noise-free template grid and noisy evaluation objects, then provided the noise is low enough, 1-NN will give exact results.  If we instead split the template and evaluation sets to have no common objects, then the precision of at least one of the APs is limited by the grid spacing.  We could instead average over the $k$ nearest neighbours, but then NN does badly because it averages over a wide range of the weak AP (a shortcoming which party motivates \ilium).
To give some comparison, however, I estimate the parameters of all 274 stars in the TG grid of noise-free spectra via leave-one-out cross validation.
The errors in both APs are six times larger than the \ilium\ result at G=15. This at least confirms that \ilium\ overcomes the grid resolution limitation of NN.

The above results are for stars with \feh\,=\,0\,dex. I retrained and evaluated \ilium\ on a grid with the full range of metallicities ($-4$ to $1.0$\,dex; TG-allmet).  The errors averaged over this full data set at G=15, reported in line 4 of Table~\ref{sumres}, are 6--8 times larger than the solar metallicity case. This is entirely due to the new metallicity variance which is not accounted for (modelled) by \ilium\ and so is a confusing factor.
(A \teff\ accuracy of 1\% at G=15 and 2\% at G=18.5 is nonethelss good for stars which a priori show variance over the full range of \teff, \logg\ and \feh).
Curiously, if we apply the model to G=20 data, then we see that the performance is no worse than when we limited the problem to solar metallicity stars (compare lines 3 and 6 of Table~\ref{sumres}). This is because the photometric noise dominates over the variance introduced by the (unmodelled) metallicity range.

If the AP residuals had a Gaussian distribution, then \rms\,=\,1.25\mar\ in Table~\ref{sumres}. The fact that \rms\ is always larger (sometimes much larger), indicates that there are outliers, which justifies the use of \mar\ as a more robust error statistic.

In terms of overall error, an SVM does better than \ilium\ on the TG-allmet problem at all three magnitudes (Table~\ref{svmres}). It's not clear why this is so, given that \ilium\ was better on the problem limited to solar metallicity (see section~\ref{sect:compare}).


\section{Application to the \teff+\feh\ problem (TM)} \label{sect:tefffeh}

I now use \ilium\ to estimate \teff\ and \feh\ on the two grids TM-dwarfs and TM-giants defined in section~\ref{sect:datasets}. In each case we effectively assume we already have a rough \logg\ estimate. The remaining spread in \logg\ in each case acts as cosmic scatter.  The models are then applied to the corresponding evaluation sets with noise levels at three magnitudes.  The summary performance statistics are show in six lines in Table~\ref{sumres}. As there is essentially no sensitivity to metallicity at \teff\ above 7000\,K (Fig.~\ref{fig:teff_feh_spectra}), I only report results on cooler stars, even though the \ilium\ models were fit to the full \teff\ range.

Looking first at the \teff\ performance, we see that the precision is twice as good as TG at all magnitudes (e.g.\ 0.007\,dex compared to 0.019\,dex).  This just indicates that we can estimate \teff\ more precisely for cool stars.  
Within the \teff\ range 4000--7000\,K there is no strong dependence of the \teff\ precision with \teff\ or \feh: \ilium\ can
estimate \teff\ equally well at all metallicities.

\begin{figure}
\begin{center}
\includegraphics[width=0.35\textwidth, angle=0]{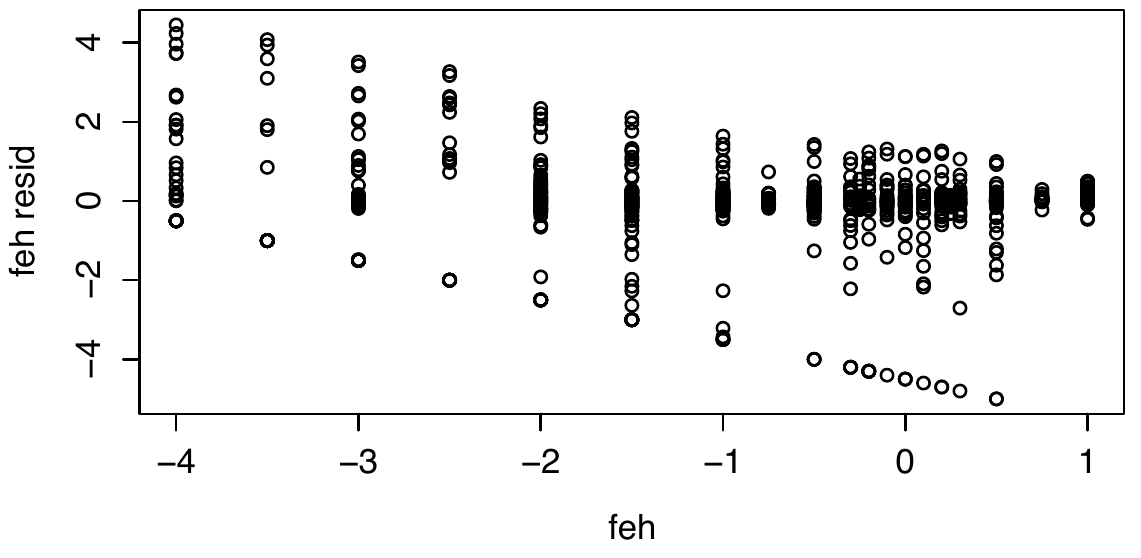}
\includegraphics[width=0.35\textwidth, angle=0]{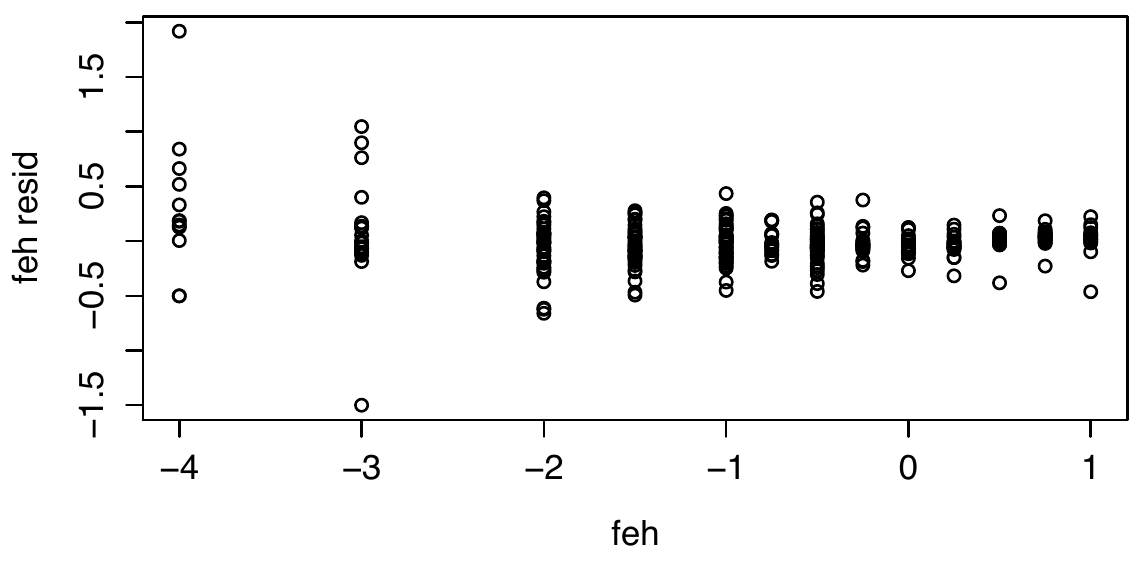}
\caption{\feh\ residuals for the TM-dwarfs model at G=15 for the full \teff\ range (top) and for cool stars only (\teff\,$\leq$\,7000\,K; bottom). Note the different scales on the ordinate. The systematic error (due to the lack of sensitivity to \feh\ for hot star spectra) vanishes in the lower panel.} \label{fig:sys_demo}
\end{center}
\end{figure}

Turning now to metallicity, we see good performance at G=15 and G=18.5 for dwarfs and giants: random errors of 0.3\,dex or less and negligible systematics. At G=20 the performance is quite a lot worse (0.7--0.8\,dex).  There is little dependence of \feh\ precision or accuracy with \feh, as can be seen in the lower panel of Fig.~\ref{fig:sys_demo}: Even for the most metal poor stars in the sample at \feh\,=\,$-$4.0\,dex the precision is still 0.5\,dex.  This plot also shows that the AP estimates hardly ever exceed the limits of the training grid, even though they are allowed to (section~\ref{sect:apextrapolate}), again suggesting natural convergence properties of the algorithm.


We gain some insight into how \ilium\ works if we include hot stars in the evaluation set. There is now a strong systematic error in the \feh\ estimates, as can seen in the upper panel of Fig.~\ref{fig:sys_demo}. The reason is that the sensitivity of all the spectral bands to metallicity is essentially zero in hot stars.  In that case there is no contribution from metallicity to the flux updates in the \ilium\ algorithm, so the flux prediction is entirely from the strong component of the forward model. That predicts a flux corresponding to the average value of the metallicity (Fig.~\ref{fig:two_component_forward_model}) in the training grid,  so this is the AP value which \ilium\ reports. This is obviously higher than the lowest metallicities, with the result that \ilium\ overestimates \feh. This is logical and desirable: \ilium\ reports the average value in the training data when the spectrum provides no information. The AP distribution is acting as a prior.

The above results have implicitly assumed the \logg\ of the star to be known well enough to identify it as a dwarf or giant. We saw from the results in the previous section that \ilium\ itself can do this (even if the metallicity is not known: line 4 of Table~\ref{sumres}). But what if we relaxed this assumption? To test this, I trained and evaluated a new model, TM-allgrav, on the full range of \logg\ (see section~\ref{sect:datasets}). Even at G=18.5 this model can estimate \feh\ to a precision of 0.4\,dex, which would be enough to trace the metallicity distribution in the Galaxy and identify very metal-poor stars.

\begin{figure}
\begin{center}
\includegraphics[width=0.45\textwidth, angle=0]{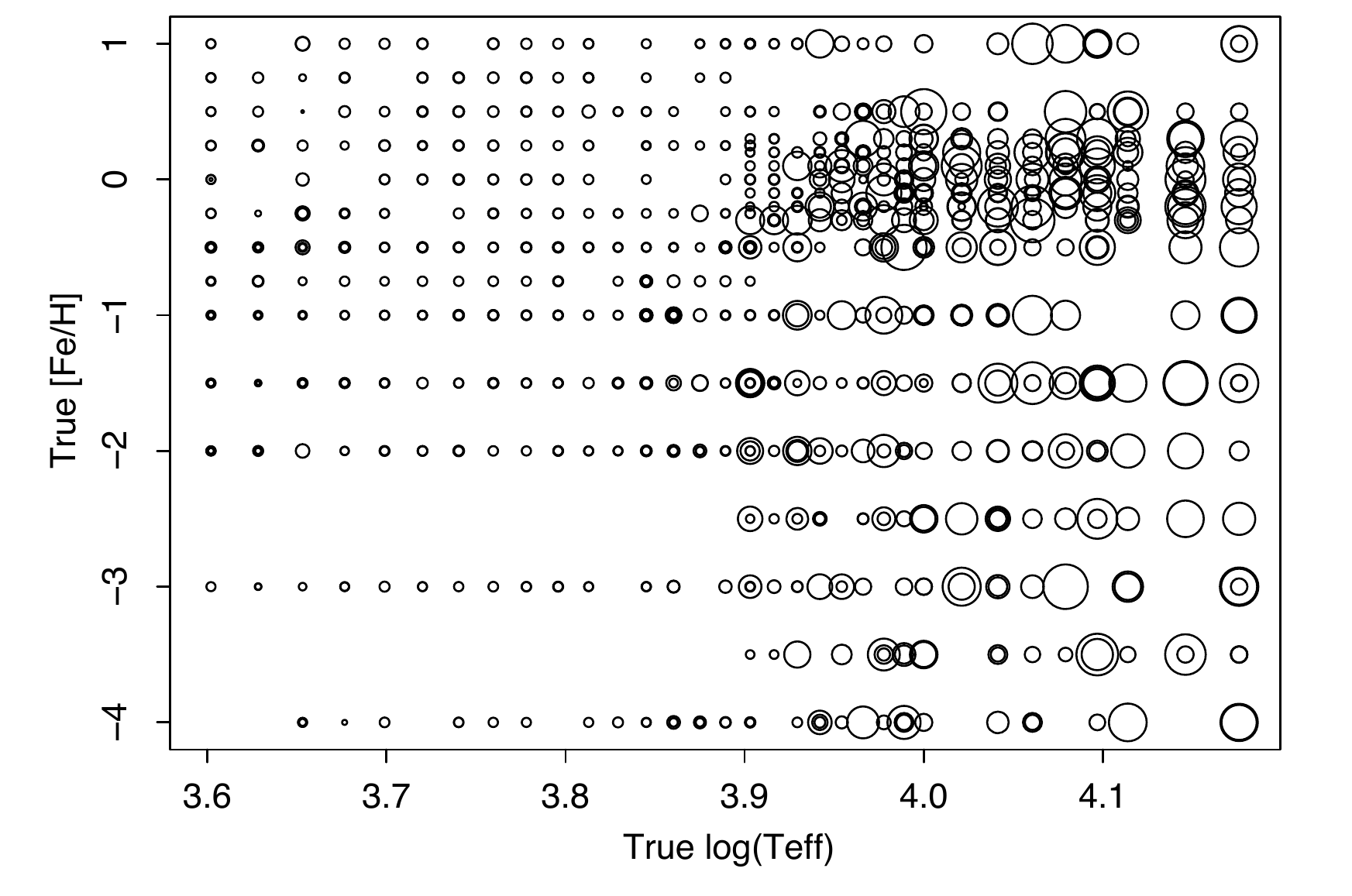}
\includegraphics[width=0.45\textwidth, angle=0]{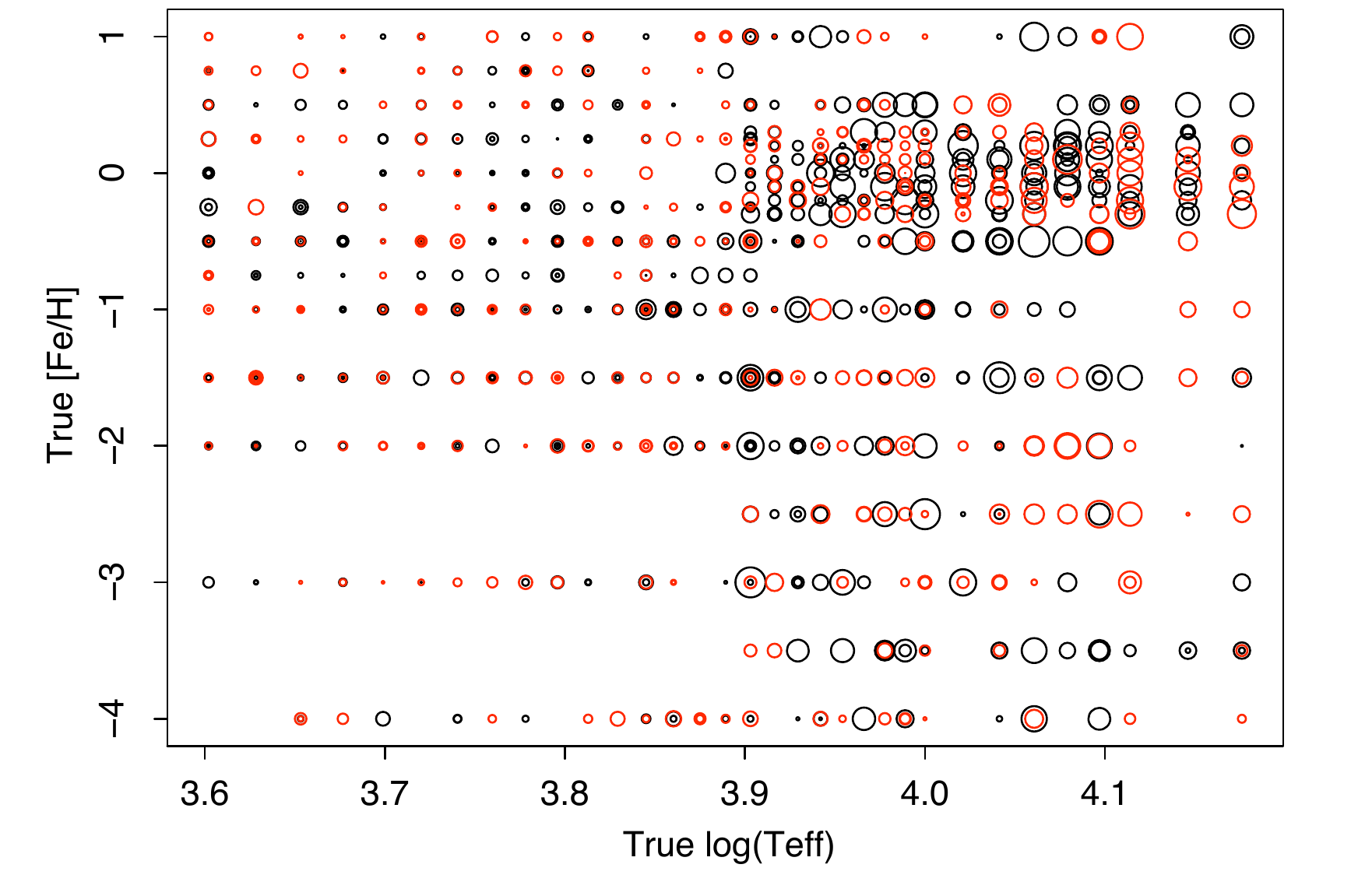}
\caption{Predicted AP uncertainties (top) and actual residuals (bottom) for \logteff\ from the TM-dwarf problem at G=20.
The area of the plotted circle (not the diameter) is proportional to the size of the uncertainty/residual. In the lower panel, positive residuals are shown in black and negative in red. Points below some size limit are not plotted at all.} \label{fig:resid_cov}
\end{center}
\end{figure}

Section~\ref{sect:errest} gave a formula for the covariance of the APs estimated by \ilium.  For this TM problem, ${\mathbfss C}_{\phi}$ is a $2\times2$ matrix with elements $c_{jj'}$. The corresponding uncertainty estimates, $\sqrt{c_{jj}}$, are plotted as a function of the two APs in the upper panel of Fig.~\ref{fig:resid_cov}. This can be compared to the actual errors (residuals) in the lower panel.  The fact that they broadly agree indicates that the uncertainty estimation is useful. This is important, because for many purposes knowing the uncertainty in an estimate is as important as the estimate itself.

\section{Application to 3-AP problems (TAG \& TAM)}\label{sect:2d1d}

\subsection{Extension to higher dimensions}

So far the forward model has comprised two 1D components. For more than two APs the forward modelling approach described in section~\ref{sect:forward_model} generalizes almost trivially. The principle is to retain the partition of the APs into the two categories ``strong'' and ``weak''.  With $N_S$ strong APs, the strong component of the forward model is a single $N_s$-dimensional function, fit at the $n_s$ unique combinations of the strong APs in the training grid.  At any one of these points, $k$, there are $n_w(k)$ grid points which vary over the $N_w$ weak APs. The mean flux over these is used to fit the strong component. The weak component at point $k$ is then built by making an $N_w$-dimensional fit to the flux residuals (with respect to the strong component) over the $n_w(k)$ points at $k$.  Thus we have $n_s$ independent weak components.  The components are combined and applied as in the 1D+1D case. Nothing else in the \ilium\ algorithm is changed.

Here I apply \ilium\ to two different problems both with two strong APs and one weak AP. This is especially important for stellar parametrization (and Gaia) because in practice we have to accommodate variable interstellar extinction, which can be large at low Galactic latitudes and has at least a large an impact as \teff\ on the spectrum (see Fig.~\ref{fig:teff_av_spectra}).

I fit the 2D strong component of the forward model using a thin plate spline, a type of smoothing spline closely related to kriging
 (e.g.\ Hastie et al.~\citealp{hastie01}).
The number of degrees of freedom is set to $n_s/2=330/2=155$ in both problems.  The 1D weak components are again fit using smoothing splines with the dof set as before (section~\ref{fmfunctions}).

\subsection{Results for (\teff+\av)+\logg\ (TAG)}

\begin{figure*}
\begin{center}
\includegraphics[width=0.70\textwidth]{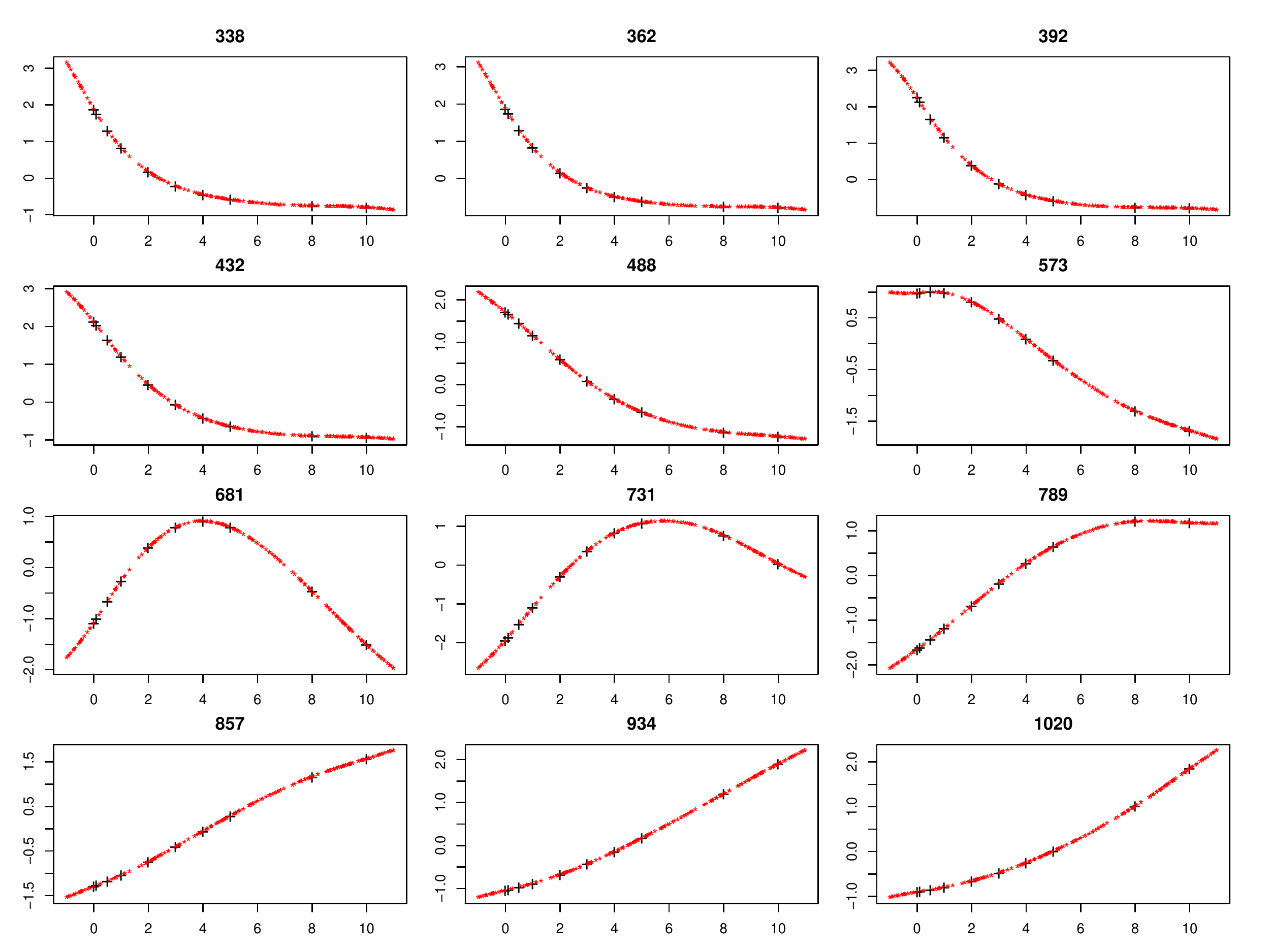}
\caption{Predictions (small red stars) of the full forward model for the TAG problem as a function of \av\ with \teff\ fixed at 10\,000\,K and \logg\ at 4.0\,dex for 12 different bands (the central wavelength of which in indicated at the top of each panel in nm). The black crosses are the (noise-free) grid points.  The flux plotted on the ordinate is in
 standardized units.}
\label{fig:tag_formod_logg=4_teff=10000}
\end{center}
\end{figure*}

In this problem the forward model is fit over the full range of \teff, \av\ and \logg\ for solar metallicity stars (TAG in section~\ref{sect:datasets}). The 1D cut through the fit in Fig.~\ref{fig:tag_formod_logg=4_teff=10000} shows that the newly introduced \av\ variance is fit accurately (as are the variations in the other APs). The summary performance when applying \ilium\ with this model to G=15 spectra is shown near the bottom of Table~\ref{sumres}. Compared to the results on the TG data set at the same magnitude, the errors in \logteff\ and \logg\ are, at 0.013\,dex (3\%) and 0.3\,dex respectively, considerably worse. This is due to the extra variance introduced by the very wide range of extinction. Yet these errors are still small enough to be scientifically useful, especially when we note that the systematics are quite small. The new AP, extinction, can be estimated very well: a mean absolute error of just 0.07\,mag. If we limit the analysis (without changing the fitted model) to just low extinction stars (\av\,$\leq1.0$\,mag), then \av\ can be estimated only slightly better (0.056\,mag) but \logteff\ more so  (0.008\,dex). 

At G=18.5, the errors of course increase (Table~\ref{sumres}): while \teff\ and \av\ are still manageable, at 1.1\,dex \logg\ is not. Fortunately Gaia will provide parallaxes for many stars to improve the \logg\ estimates (given the \teff\ and \av\ estimates from \ilium).

At G=15 the SVM has comparable performance on this problem, but is somewhat better at G=18.5 (see section~\ref{sect:compare}).

\subsection{Results for (\teff+\av)+\feh\ (TAM)}

We now swap \logg\ for \feh\ and train and evaluate \ilium\ on the TAM problem (which is a dwarf sample: the results are broadly similar for a giant sample).
The summary performance at G=15 is listed near the bottom of Table~\ref{sumres}. The performance is degraded significantly compared to the case with no \av\ variance (TM-dwarfs), much more so for \teff\ than for \feh. At 0.46\,dex the metallicity error is acceptable. Alas at G=18.5 this degrades to an almost useless level with the additional problem of a large systematic.\footnote{Contrary to expectations, systematic trends can rarely be corrected for. We would have to plot the residual vs.\ the {\em estimated} AP, and if the scatter is larger than the systematic trend then a correction cannot be made.}

If we limit the analysis (at G=15) to just low extinction stars (\av\,$\leq1.0$\,mag), then the \feh\ precision is essentially unchanged (0.44\,dex). This implies that it is no more difficult (on average) to determine the metallicity of stars with high extinction than of stars with low extinction.  
So why is the error larger here than in the zero extinction case reported for the TM problem? It is because we now have a large extinction range {\em a priori}, so an uncertainty in determining \av\ corresponds to an uncertainty in \feh. As \feh\ is weak, a relatively small uncertainty in \av\ is magnified into a larger one in \feh.

\begin{figure}
\begin{center}
\includegraphics[width=0.5\textwidth]{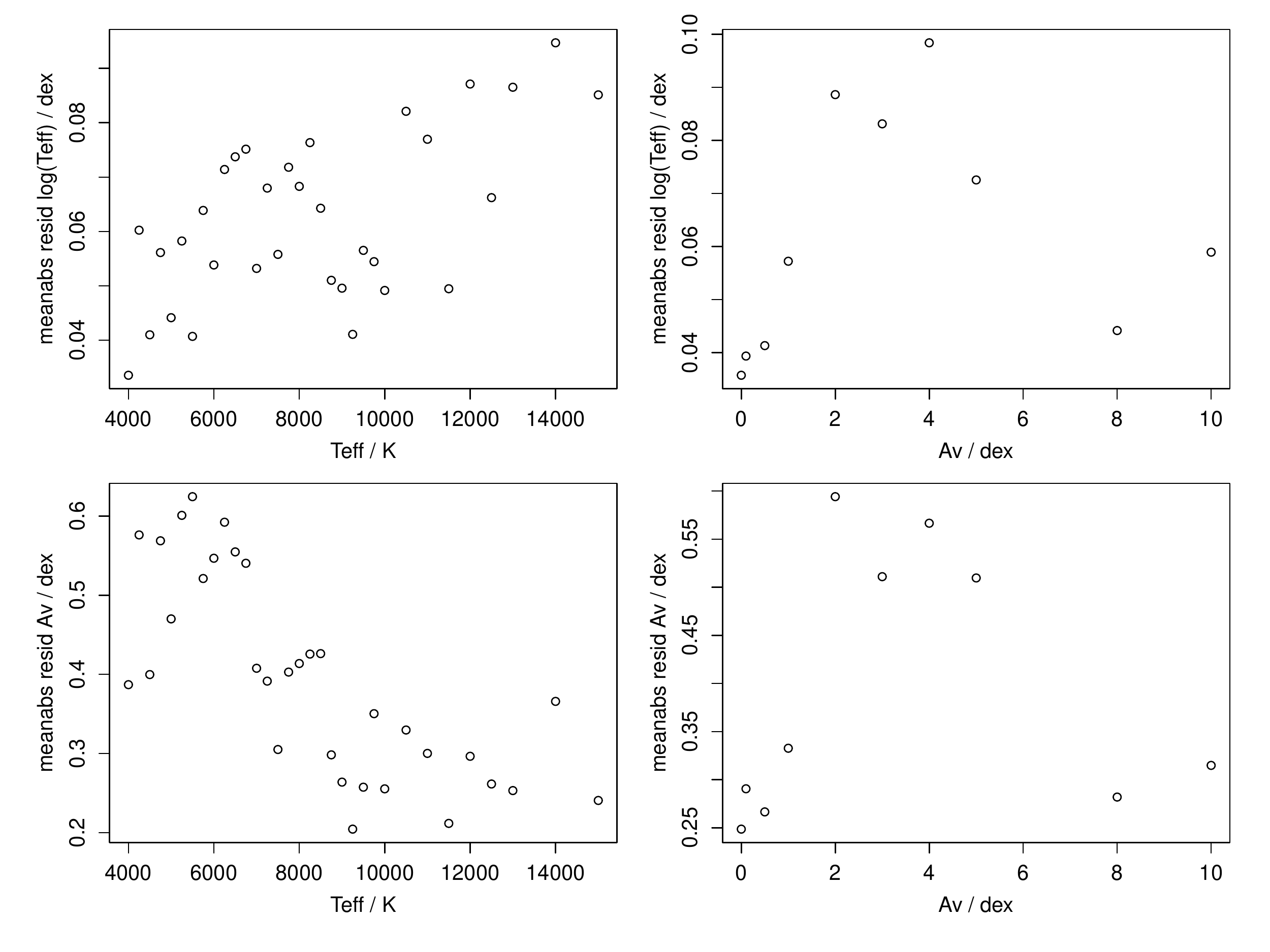}
\caption{The trend in the mean absolute errors
for \logteff\ and \av\ in the TAM problem on the G=18.5 evaluation data set
(which spans the full range of \teff, \av\ and \feh)}
\label{fig:tam_residtrend}
\end{center}
\end{figure}
 
Note that the extinction errors are larger than those reported with the TAG problem (0.18\,mag compared to 0.07\,mag at G=15).  This is a consequence of having limited the analysis to cool stars. Indeed, the mean absolute error on the present problem is only 0.1\,mag if we analyse the hot stars (\teff\,$>$\,7000\,K) rather than the cool ones:
\av\ can be estimated more accurately for hotter stars.
This is not surprising because hot star spectra are simpler (e.g.\ no metallicity signature) so it should be easier to untangle the effects of temperature and extinction. This is illustrated in Fig.~\ref{fig:tam_residtrend} at G=18.5, which plots the dependence of the \logteff\ and \av\ residuals with these parameters (averaged over all \feh). It also shows that \teff\ is estimated more precisely for cool rather than hot stars, as found also for the TM problem.

\begin{figure}
\begin{center}
\includegraphics[width=0.30\textwidth]{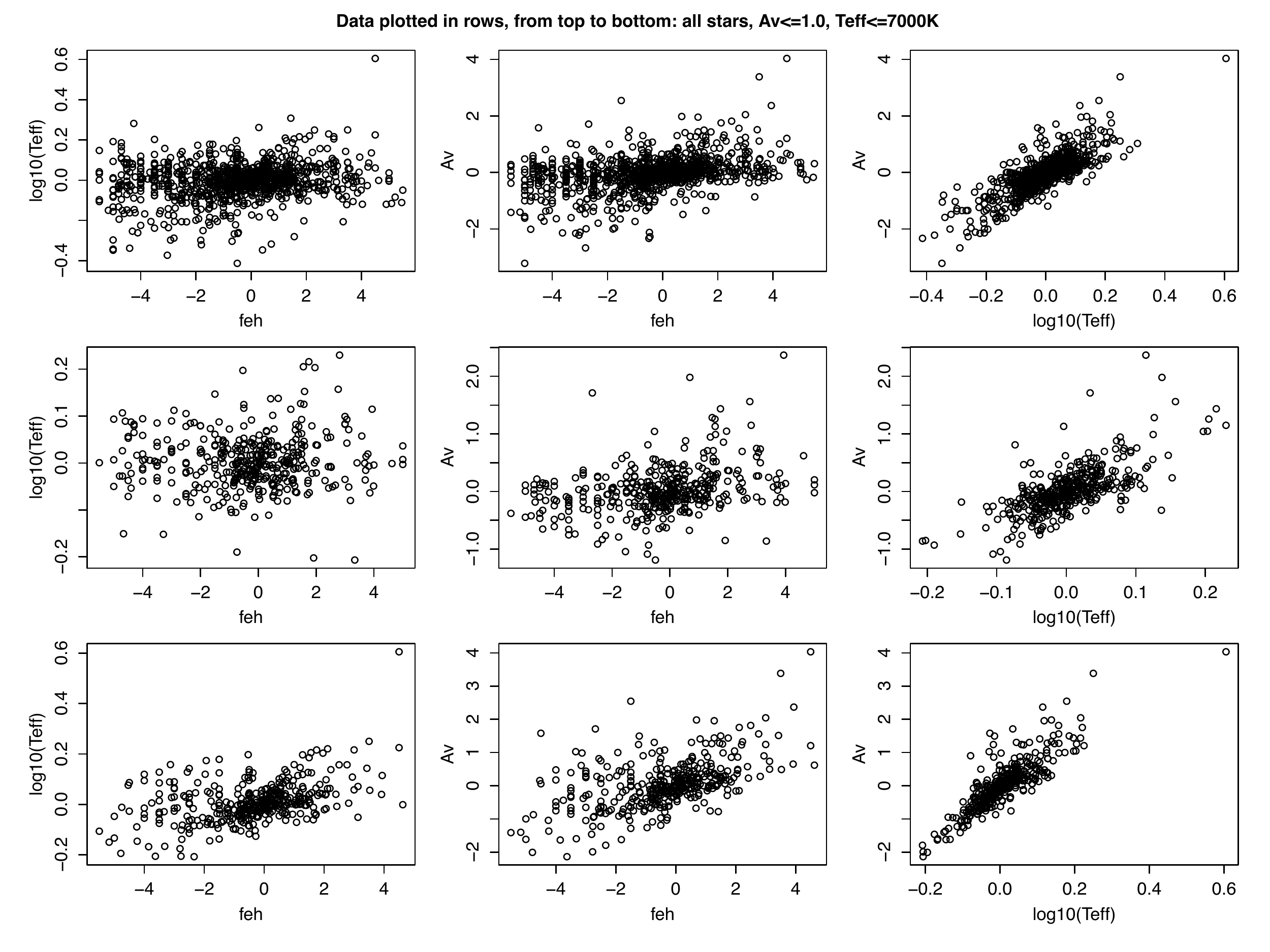}
\caption{The correlation between the \av\ and \logteff\ residuals for the TAM model applied to the
G=18.5 evaluation data set}
\label{fig:tam_resid_correlation}
\end{center}
\end{figure}
 
Nonetheless, the \teff\ and \av\ errors are quite large at G=18.5: 8--20\% for \teff\ and 0.2--0.6\,mag in \av\ (Fig.~\ref{fig:tam_residtrend}). Moreover, their residuals show a strong positive correlation (Fig.~\ref{fig:tam_resid_correlation}): a tendency to overestimate one results in an overestimation of the other.  This suggests that these two APs are degenerate in these low resolution BP/RP spectra, something which Fig.~\ref{fig:teff_av_spectra} also suggests. Let us now investigate this further.

\subsection{Identification of a \teff--\av\ degeneracy}\label{degeneracy}

Like most estimation algorithms, \ilium\ simply tries to find the best single solution (plus an uncertainty estimate).  If we consider a likelihood function of the data given the APs, $P({\rm Data}|{\bmath \phi})$, then the algorithm is trying to find the maximum of this, plus some measure of the width of this peak.  But if there is a degeneracy in the APs then the peak position and width are an inadequate summary of the likelihood.
A degeneracy can be defined as two or more stars which have spectra differing by an amount consistent with the photometric noise. On observing one of these spectra we would be unable to distinguish between the AP solutions. This applies trivially to stars with very small AP differences (thus the need to quote an uncertainty). What we are interested in here are objects which have significantly different APs (a large fraction of the grid range), such that a simple approximation of the likelihood function (e.g.\ a Gaussian in the APs with covariance given by equation~\ref{eqn:apcov}) fails to represent the uncertainties accurately.

\begin{figure}
\begin{center}
\includegraphics[width=0.4\textwidth]{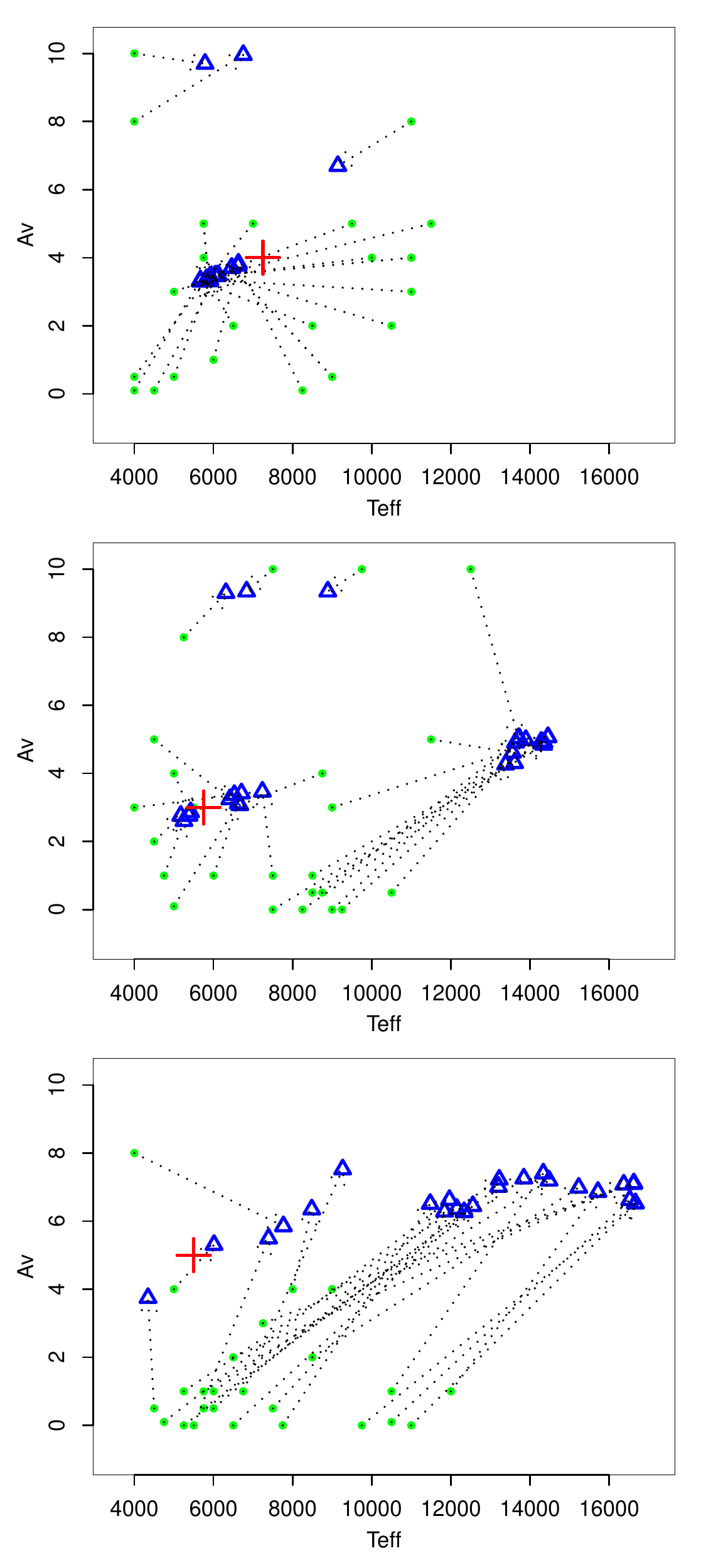}
\caption{Multiple random initialization of \ilium\ for three different stars (in the three panels).
In each panel the starting points are shown as green dots connected to the solutions (after 20 iterations) shown as blue triangles. The true APs are shown as a red cross.}
\label{fig:multi_init}
\end{center}
\end{figure}
 
As \ilium\ is a local search method, it is a priori possible that when initialized at different points it could converge on different (degenerate) solutions.  To explore this I ran \ilium\ 25 times for each star with the initialization chosen at random each time, rather than by the nearest neighbour. (I used the TAG model with G=18.5 spectra.) Fig.~\ref{fig:multi_init} plots the various solutions in the \teff--\av\ plane for three different stars, chosen to illustrate degeneracy.

For the first star (top panel), we see that 22 of 25 initializations over quite a range of \teff\ and \av\ converge on a small region (slightly offset from the true solution possibly due to noise in the spectrum).  The three ``wrong'' solutions occur because the algorithm is initialized too far from a good solution: the searching method gets stuck in a poor local minimum. Inspection of their predicted spectra shows them to be very different from the true spectrum, with GoF values (equation~\ref{eqn:gof_chisq}) of more than a few hundred. Such a high GoF would be used in practice to flag and reject such solutions.

Turning to the second star (middle panel) we see two distinct regions of convergence, one of which is close to the true APs.  Are the other solutions at around \teff\,=\,13\,000\,K simply poor (bad convergence)? When we inspect the predicted spectra (not shown), we see this is not the case. The two sets of spectra are indistinguishable with the noise, all having similar and small GoF values (0.01--15).

\begin{figure*}
\begin{center}
\includegraphics[width=1.0\textwidth]{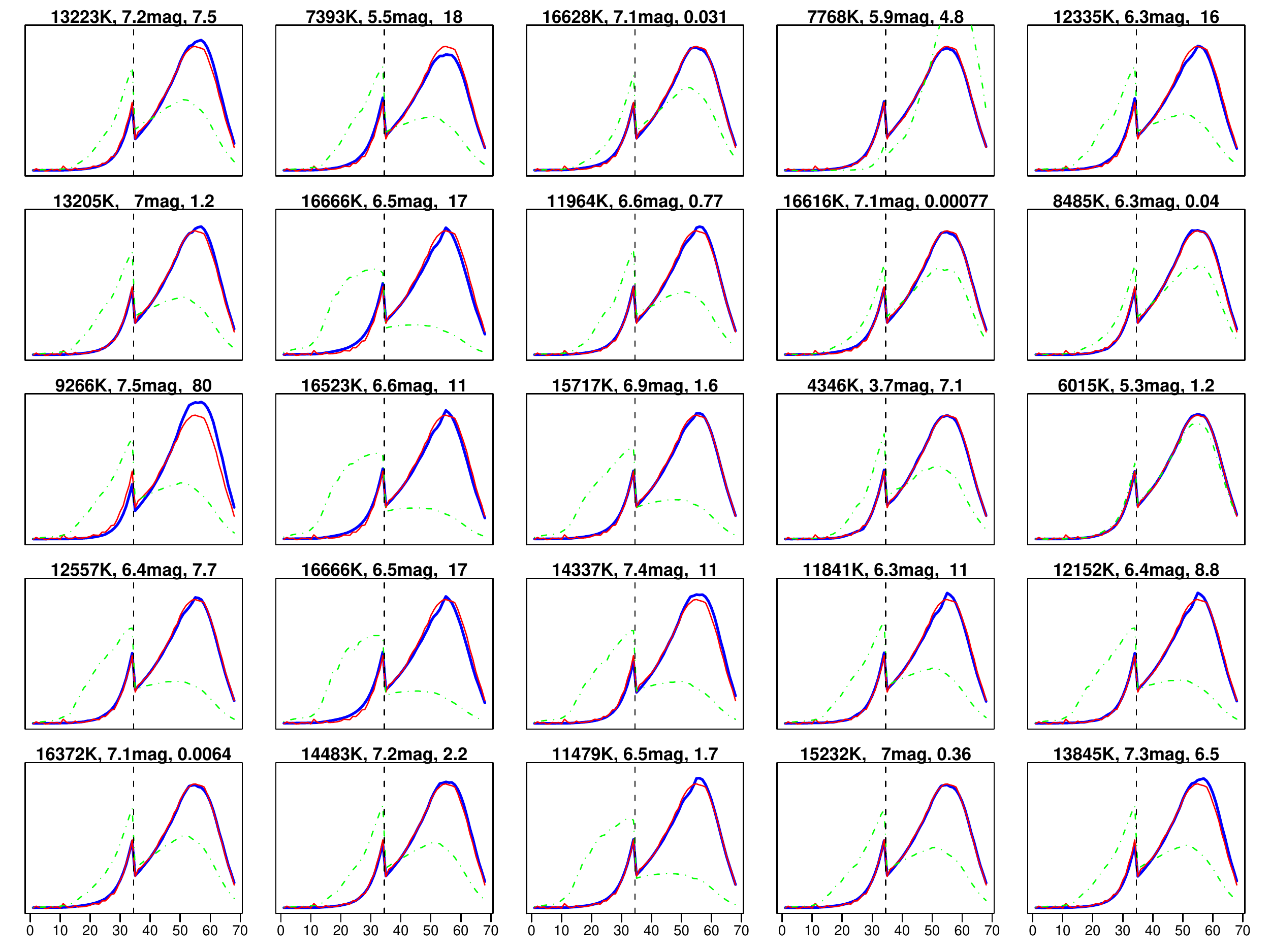}
\caption{The initial (green dashed-dot line), final (blue thick line) and true (red thin line) spectra for the 25 runs of \ilium\ for the one star shown in the bottom panel of Fig.~\ref{fig:multi_init}, plotted vs.\ band number (BP and RP are separated by the vertical dashed line). The \teff\ and \av\ of the \ilium\ solutions (the final spectra) are plotted at the top of each panel along with the GoF of each. The true APs are
5500\,K and \av\,=\,5.0\,dex}
\label{fig:photpred}
\end{center}
\end{figure*}

This is better illustrated by the third star (bottom panel), where we see a lack of convergence on one or two solutions in favour of a complete ridge of solutions. Fig.~\ref{fig:photpred} plots the initial and final (predicted) spectra for these 25 runs of \ilium. All but one or two of these final spectra agree very closely with the true spectrum, even though they have very different APs.
So this is a true degeneracy, and not poor convergence of \ilium.  This degeneracy is also reflected by
a high value for the expected correlation coefficient between \teff\ and \av\ calculated by equation~\ref{eqn:apcov}.  Identification of similar ridges in AP space of near-identical spectra for other stars suggests that the \teff--\av\ degeneracy is widespread.

\subsection{Mapping of the \teff--\av\ degeneracy}\label{degeneracy2}

\begin{figure*}
\begin{center}
\includegraphics[width=1.0\textwidth]{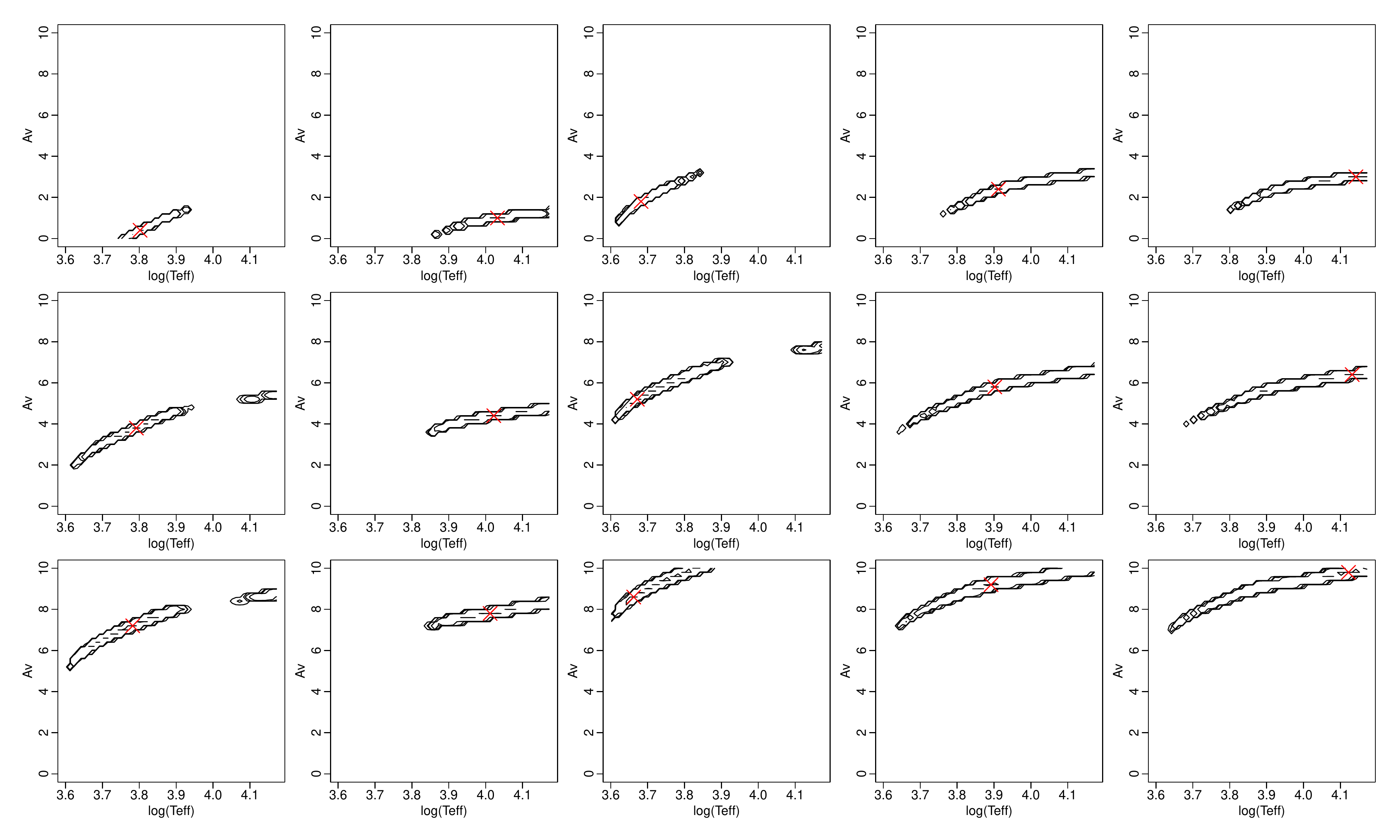}
\caption{Degeneracy map (a likelihood function) for the forward model of the TAG grid at G=18.5. For each of the 15 stars (panels) marked as a red cross, each contour indicates those stars which have a probability $P$ of having a spectrum indistinguishable to within the expected photometric noise. Contours are plotted for $logP = \{-4, -3, -2, -1, 0\}$ (due to the steep sides of the \chisq\ function, these are barely distinguishable).}
\label{fig:dgenmap}
\end{center}
\end{figure*}

We can map this degeneracy systematically using just the forward model to generate spectra
on a fine grid of \teff\ and \av. For each predicted spectrum I adopt as its sigma spectrum 
that of the star in the original grid which has the closest APs.
The expected Mahalanobis distance between a hypothetically measured spectrum,
${\bmath p}$, and any noise-free spectrum, ${\bmath p'}$, in this grid is, with $\delta{\bmath p} = {\bmath p} - {\bmath p}'$,
\begin{eqnarray}\label{eqn:dist}
D^2 \, &=& \, \delta{\bmath p}^T {\mathbfss C}_p^{-1} \delta{\bmath p} \\ 
           &=& \, \sum_{i=1}^{i=I} \left ( \frac{p_i - p'_i}{\sigma_{p_i}}  \right )^2
\end{eqnarray}
the simplification following because in these simulations there is no inter-pixel noise correlation. 
A degeneracy arises between two stars when $D^2$ is sufficiently small that it could arise just from photometric noise.  Under the null hypothesis ($H_0$) that the differences between the spectra are only due to Gaussian noise and that each pixel is independent, $D^2$ follows a \chisq\ distribution with $I-1=67$ degrees of freedom. I will define two stars as non-degenerate only if the probability of observing their given distance or more under $H_0$ is 1\% or less. 
This corresponds to a critical value of $D^2_{\rm lim}= 96.8$ (or a reduced-\chisq\ value of 1.44). In other words, stars separated by a smaller distance are considered degenerate.\footnote{\chisq\ is the distribution followed by a sum of squares of independent unit Gaussian variables, $N(0,1)$, and has a (unnormalized) density function
$P'(D^2) = D^{\nu-2} e^{-D^2/2}$ where $\nu$ is the degrees-of-freedom. $\nu\!=\!I\!-\!1$ because one degree-of-freedom is ``lost'' by the fact that they have a common G magnitude, although this is of no practical significance here. 
$D^2_{\rm lim}$ is defined by
$P(D^2 \geq D^2_{\rm lim} | H_0) = \int_{D_{\rm lim}^2}^{\infty} \! P' dD^2$ = 0.01.
Fig.~\ref{fig:dgenmap} plots this probability (and several others) as a contour.} 

For each star in the fine grid I calculate the distance to all of the other stars.  The corresponding probabilities (of getting that distance or more) can then be plotted as a {\em degeneracy map} in the \teff--\av\ plane.  Maps for 15 stars are shown in Fig.~\ref{fig:dgenmap} (similar patterns are observed throughout the grid).  We see that the APs are correlated, the high probabilty region extends over a large range of the grid and in some cases is multimodal.  In other words, there is a strong degeneracy over the whole plane.  What this means in practice is that if we observed a (noisy) spectrum, ${\bmath p}$, at the position of one of the crosses in Fig.~\ref{fig:dgenmap}, then this spectrum is indistinguishable from all the spectra lying within the contours, on account of the noise.  Therefore we also cannot distinguish between the corresponding APs.

We can compare the degeneracy maps with the expected covariance for individual objects calculated from equation~\ref{eqn:apcov} by treating the latter as the covariance in a Gaussian likelihood function in the APs. While the estimated correlations agree qualitatively with what we observe in Fig.~\ref{fig:dgenmap}, this Gaussian likelihood approximation does not accurately reproduce the shape of the degeneracy ridges. Indeed it cannot, because the degeneracy ridges are not symmetric about the true estimate, and \av\ cannot be negative. This lack of detailed agreement is not very surprising, however, because equation~\ref{eqn:apcov} is a valid approximation only when the linear relation in equation~\ref{eqn:deltap} holds, i.e.\ when the errors are small.

The implication of this is significant: It is misleading to report a single estimate for \teff\ and \av\ (even if accompanied by a simple covariance estimate). Rather, we must report a whole ridge of solutions. As we can map these degeneracies in advance, the purpose of a classifier such as \ilium\ would be to identify {\em a} solution, which we use to identify the corresponding degeneracy ridge (either as a probability grid or a fitted approximation to it).  This study shows that a single nearest neighbour initialization of \ilium\ is usually adequate to allow \ilium\ to find a good solution and therefore the correct ridge. The minority of cases where no good solution is found are usually identifiable by the large GoF value. In those cases we may want to re-initialize \ilium\ from some other point or improve the convergence mechanism.

Note that Fig.~\ref{fig:dgenmap} is for G=18.5. At G=20 the degeneracy regions are somewhat larger.  At G=15 the SNR is sufficiently high that there is no longer any degeneracy on a scale of practical relevance: the contours in the degeneracy map are smaller than the grid point separation (0.2\,mag in \av\ and 0.01\,dex in \logteff). This is consistent with the small errors in these APs recorded in Table~\ref{sumres}.

\subsection{Including additional or prior information}

Additional information could help to reduce the extent of the degeneracy. 

The parallax reported by Gaia combined with the apparent G-band magnitude gives an estimate of \mg$+$\ag, where \mg\ is the absolute magnitude in the $G$ band and \ag\ is the extinction in the G-band (which we could estimate directly from \ilium\ instead of \av). Combined with an HR-diagram (which is just a prior probability density function over \teff\ and \mg) this provides additional constraints on \teff\ and \ag. A method for doing this is outlined in Bailer-Jones~\cite{cbj10}.

We could set weak prior probabilities on the APs -- the probability of the APs unconditional on the BP/RP spectrum, parallax or G-band magnitude -- based on a simple model of the Galaxy. For example, we know that cool stars are much more common than hot stars, so an unreddened cool star is a priori more likely that a highly extinct OB star.  Likewise, if the star is observed at high Galactic latitude we could confidently assign a lower prior probability to high extinctions.  This latter prior is particularly useful, because we saw in sections~\ref{sect:tefflogg} and \ref{sect:tefffeh} that if the extinction can be fixed then \ilium\ gives much better estimates.  Normally, however, we wouldn't want to bias the Gaia AP estimates with {\em detailed, current} knowledge of Galactic structure, given that the primary goal of Gaia is to improve this knowledge.

For very bright stars (G$\ltsim$12; Gaia will saturate at around G=6) we may have AP estimates coming from the high-resolution RVS spectrograph on Gaia measuring the CaII triplet at 860\,nm (Katz et al.~\citealp{katz04}). In principle we can estimate \teff\ independently of \av\ using these data. Note that because the ridges in Fig.~\ref{fig:dgenmap} have a low inclination, additional information on \teff\ is more useful than information on \av.

Finally, estimates of APs from other surveys could be incorporated and combined with the Gaia ones, provided they can be interpreted as probability distributions.

Given the opportunity, one should design an ad hoc spectrophotometric system to address the specific goals of the survey. Particularly desirable is
a system which maximizes the separability of the APs (a method for doing this was presented by Bailer-Jones~\citealp{cbj04}).
This opportunity was taken by the Gaia community, which designed a multiband photometric system to address the specific Gaia scientific goals (Jordi et al.~\citealp{jordi06}). But this was later replaced with the present (and less desirable) BP/RP spectrophotometry, ultimately for financial reasons.

\section{Further applications (and the TGM problem)}\label{sect:more}

In the previous section the forward model was extended to be 2D in the strong APs and 1D in the weak.  In a similar way it can be adapted to model one strong AP (e.g.\ \teff) and two weak APs (e.g.\ \feh\ and \logg) simultaneously. This is of practical use in situations where extinction can be assumed to be low.  I used this to fit a forward model to the TGM grid (section~\ref{sect:datasets}) and applied \ilium\ as before. The results are shown in the bottom section of Table~\ref{sumres}. At G=15, the performance in \teff\ and \feh\ is considerably better than obtaind with the corresponding TM problems in which there was unmodelled \logg\ variation: we now obtain 0.3\% in \teff\ for the full \teff\ range and 0.1\,dex in \feh\ for cool stars. \logg\ is also better than found with TG-allmet, but not with TG where metallicity was fixed to zero.  This as we would expect: at high SNR, modelling the extra variance improves the results in all APs.

This is not the case at G=18.5. Here the performance of the TGM model on all three parameters is similar to that obtained in the TG-allmet or TG-allgrav problems on the appropriate \teff\ ranges. I suspect that the extra variance introduced by the noise is complicating the problem, so that the modelling the additional AP (rather than marginalizing over it) brings no improvement in performance (but the obvious advantage that we now determine all three APs instead of just two.)  It is also worth pointing out that when using the degeneracy mapping method described in section~\ref{degeneracy2}, I find there is a very strong and complex degeneracy between \feh\ and \logg\ at G=18.5. 

\ilium\ could be further extended to more or other APs or to different problems, provided the partition between strong and weak APs can be retained. If such a clear distinction were not possible then a different approach to forward modelling would be necessary. If we had more than two or three weak or strong APs, then the smoothing splines are unlikely to provide adequate fits, unless we had a lot of data. To overcome this we would need more structured regression models suited for fitting high dimensional sparse data sets (e.g.\ neural networks).

\section{Summary and Conclusions} \label{conclusions}

I have introduced an algorithm for estimating parameters from multi-dimensional data.  It uses a forward model of the data to effectively perform a nonlinear interpolation of a template grid via the Newton-Raphson method.  It is intended to overcome both the non-uniqueness issue of direct modelling of inverse problems as well as the finite grid density and metric definition issues of $k$ nearest neighbours. \ilium\ makes use of the sensitivity of the data to the astrophysical parameters to find an optimal solution.  This is convenient, because in principle it means we don't need to worry too much about feature selection: low sensitivity features will automatically be down weighted.
An important component of the algorithm is the division of the forward model into two parts which independently model the variance of the ``strong'' APs (such as \teff\ and \av) and the weak APs (such as \logg\ and \feh). Good fits could be obtained using low-dimensional (1 or 2) smoothing splines.  As it is based on a forward model, \ilium\ naturally provides AP uncertainty estimates (actually full covariances) and a goodness-of-fit, in contrast to most inverse modelling methods.

I applied \ilium\ to the problem of estimating APs from simulations of the low resolution Gaia spectrophotometry, BP/RP. Results are summarized in Table~\ref{sumres}.  When limited to zero extinction stars, \teff\ can be estimated at G=15 to a (mean absolute) accuracy of better than 1\% when the metallicity and surface gravity are entirely unknown
(\logg\,=$-0.5$ to $5.0$\,dex; \feh\,=\,$-4$ to $+1.0$\,dex).
At G=18.5 and G=20.0 (the limiting Gaia magnitude) the average \teff\ accuracy over the full range of APs is 2\%
and 4\% respectively.
At G=15 \logg\ can be estimated to 0.5\,dex if \feh\ is entirely unknown but to 0.15\,dex if \feh\ is also modelled (full  range of the three APs). Limiting to solar metallicity stars this improves to better than 0.1\,dex and is still 0.35\,dex at G=18.5.  \feh\ for cool stars can be estimated to 0.1\,dex at G=15 and to 0.15--0.35\,dex at G=18.5, the better result obtained if we know the star is a dwarf.  At G=20 \logg\ and \feh\ cannot be estimated to any useful accuracy.  Of course, for population studies we can average over many stars to reduce the population estimate of metallicity, limited by the systematic errors and any correlations in the data.  

If we extend the strong component of the forward model to simultaneously model \av\ over a very wide range (0--10\,mag), then the performance on the weak APs is significantly degraded on account of this extra variance, such that reasonable accuracies can be reached at G=15 but not at G=18.5. \av\ itself can be estimated remarkably well, however, 0.07--0.2\,mag at G=15 and 0.3--0.5\,mag at G=18.5. The extra variance also affects the \teff\ determination. However, these statistical errors do not tell the whole story: I have shown that there is a strong and ubiquitous degeneracy between \teff\ and \av\ intrinsic to the BP/RP spectra. Thus in addition to reporting single optimal estimates, the Gaia catalogue will need to provide the corresponding degeneracy map, which can be built in advance using the forward model. The degeneracies could be reduced (and the weak APs also then estimated more accurately) if we can use additional information. Possible sources include the apparent magnitude and parallax measured by Gaia, external data, and/or weak priors from a very simple Galaxy and stellar evolution model.

For some problems \ilium\ outperformed a support vector machine in terms of smaller residuals, but in other cases the SVM was better.  This should be assessed in more detail after improvements to the \ilium\ algorithm have been explored. For example, the updating method is very simple -- stopping after a fixed number of iterations -- whereas a more adaptive approach may help on larger variance problems (e.g.\ noisier data). We may also want the AP update weighting to take into account the noise (and not just the sensitivity). 

Now that we have a forward model, further possibilities for modelling open up. \ilium\ is just a method for locating the best APs. If we define a distance metric (such as equation~\ref{eqn:dist}), then we can define a likelihood function, $P(D|{\bmath \phi})$, which when combined with a suitable prior defines a (non-analytic) posterior over the APs, $P({\bmath \phi} | D)$.
We can then use one of many sampling methods, such as Markov Chain Monte Carlo, to sample this as a function of ${\bmath \phi}$, thus yielding a complete Bayesian probabilistic solution which varies smoothly over the APs.  This approach to parameter estimation has been used in several areas of astronomy, such as galaxy classification (e.g.\ Heavens et al.~\citealp{heavens00}) and inference of cosmological parameters (e.g.\ Percival et al.~\citealp{percival07}).  (If we dispensed with the forward model and just calculated probabilities at points in the original grid, then the principle is similar to that used by Shkedy et al.~\citealp{shkedy07}.)  While offering some advantages, this approach is much slower than \ilium, because for the present application it would require of order $10^4$ to $10^5$ samples and hence this many evaluations of the forward model, compared to about $10^2$ for \ilium.


\section*{Acknowledgements}

I would like to thank Anthony Brown, Lennart Lindegren and Carola Tiede for useful discussions.
This work makes use of Gaia simulated observations, which have been produced thanks to the efforts of many people
in the Gaia DPAC. In this respect I would particularly like to thank Anthony Brown, Yago Isasi, Xavier Luri, Paola Sartoretti, Rosanna Sordo and Antonella Vallenari, without whose efforts the predictions of the Gaia performance would not have been possible. The GOG simulations were produced using the MareNostrum supercomputer at the Barcelona Supercomputing Center -- Centro Nacional de Supercomputaci\'on. I am also grateful to the Marcs team at Uppsala University for producing new stellar spectral simulations for Gaia data processing purposes.


\appendix

\section{Assessing model performance}\label{sect:modelassess}
 
The \ilium\ forward model should obviously be fit on the full available grid and using noise-free data. (We need to achieve a minimum grid density in order to reliably assume that the modelled signal vary smoothly between the grid points, yet the grid need be no denser. Some smoothness assumption must be made by any method of estimating continuous parameters.)  Yet at first sight one may consider it illegitimate to fit \ilium\ on the full grid and then assess its performance on (noisy) spectra selected from the same grid. (Note that the initializaton set is never included in the evaluation set!) The equivalent approach is generally considered invalid for inverse models, because these are fit by minimizing the AP error on a training set: If they are not properly regularized the model will overfit the training data by learning non-general aspects of these data (such as the noise).
But the situation with \ilium\ is different, because the best forward model fit is identified (interactively at present) according to its smoothness and the photometry error, not the AP error. That is, the regularization is also done independently of the AP error.
It might anyway seem preferable to start with a grid twice as dense as required, fit the forward model on half the data and evaluate \ilium\ on the other half. But if the smoothness requirement is met, this will anyway give a similar performance.

The underlying issue here -- relevant to all machine learning methods -- is how to make a fair assessment of performance. Ultimately all assessments are limited by the fact that training and evaluation data are drawn from a common grid. Yet it would be useless to train the classifier on spectra generated by one set of stellar models and evaluate it on spectra generated by another, because this ``performance'' would reflect the intrinsic differences between the stellar models.  This remains a quandary, also because tests based on synthetic spectra ignore the realities of non-Gaussian noise and cosmic scatter of real data.  The most reliable assessment of performance would be to first determine APs of a set of high resolution spectra (using whatever method) based on a particular stellar model and define these APs as true. We would then degrade the spectra to the dispersion, line-spread-function and noise of real Gaia data and compare the \ilium\ estimates with the true ones. Such tests are planned but are considerably more arduous than what's done here.

\section{Comparative performance of a support vector machine}\label{sect:compare}

It is not the goal of this article to make a detailed comparison between the accuracy of \ilium\ and conventional machine learning methods, but a brief comparison is useful. I therefore applied a support vector machine (SVM) (e.g.\ Cortes \& Vapnik~\citealp{cortes95}, Burges~\citealp{burges98}) to some of the problems presented in the article. The SVM is used to directly model the inverse problem by training on noisy data.  Although the SVM training has a unique solution for a given set of data, it has three hyperparameters (the length scale parameter $\gamma$ and two regularization parameters $\epsilon$ and $C$) which must be optimized (``tuned''), which is just a higher-level training procedure. 
I optimize these hyperparameters simply via a brute force search of a regular grid over the hyperparameters, typically with 512 or 1000 models (8 or 10 values of each hyperparameter).  To achieve this the data set must be split into three independent parts: the training set, a test set (used to select the best combination of hyperparameters) and the evaluation set (on which the final performance is calculated).
(If we just use the same data for testing and evaluation -- as is frequently done -- then the results are somewhat better; unfairly so, because then the SVM is tuned on the very data set on which it is finally evaluated. I nonetheless did this for the TG problem because of the small amount of data: just 274 stars.) For each of the problems/data sets described in section~\ref{sect:datasets}, the set is randomly split into three, equal-sized disjoint parts to build these subsets. The SVM is tuned on noisy data, separately for each magnitude, and separately for each AP (as an SVM can only model one output).

\begin{table}
\begin{center}
\caption{Performance of an SVM on a selection of the problems reported in Table~\ref{sumres}. The error statistic is the mean absolute error, \mar. Figures in bold indicate that the error is significantly larger (by more than a third) than the \ilium\ error, and vice versa for numbers in italics. Note that the models for the TG problem used just a two-way split into train/test sets due to lack of data, which gives them optimistic results.}\label{svmres}
\begin{tabular}{ll*{4}{r}}
\hline
model & test sample         & \av\        & \logteff\        & \logg\        & \feh\ \\
\hline
TG & F G=15                    &                & {\bf 0.0015} & {\bf 0.10}    & \\
TG & F G=18.5                 &                & 0.0064          & {\bf 0.47}    & \\
TG & F G=20                    &                & 0.016            &  0.90     & \\
TG-allmet & F G=15         &               & {\em 0.0017} & {\em 0.091} & \\
TG-allmet & F G=18.5      &               & {\em 0.0060} &  {\em 0.41}  & \\  
TG-allmet & F G=20         &               & {\em 0.013}   & {\em 0.77}    & \\
TM-dwarfs & L G=15        &                & 0.0019          &                     & {\bf 0.20}    \\
TM-dwarfs & L G=18.5     &                & {\bf 0.0064}  &                    & {\bf 0.37}    \\
TM-dwarfs & L G=20        &                &  {\bf 0.013}   &                    & 0.63    \\
TAG & F G=15                  & 0.084      &  0.015           &   0.029  & \\
TAG & F G=18.5               & 0.24        &  {\em 0.045}  &   {\em 0.62} \\    
\hline
\end{tabular}
\end{center}
\end{table}

The SVM results are summarized in Table~\ref{svmres}. This obviously conceals details, such as the existence of systematic errors in some cases. For example, \teff\ is systematically underestimated for \teff\,$\geq$\,10\,000\,K and \logg\ shows a systematic across the whole \logg\ range in the TAG problem at G=18.5. Comparing with the \ilium\ results in Table~\ref{sumres}, no very clear pattern emerges, with \ilium\ superior to SVM in some problems and vice versa in others.  When comparing performance star-by-star on a given problem, we do not see that one algorithm performs systematically better than the other in some parts of the AP space. Rather, one algorithm is just overall better.
The fact that \ilium\ is superior on the TG problems whereas SVM is superior on the TG-allmet problems suggests that SVM copes better with problems where there is more unmodelled variance, although the better performance of \ilium\ on the TM-dwarfs problem and TAG at G=15 does not support this suggestion.  Possibly it is when the variance is higher overall (lower SNR and/or unmodelled variance) that SVM gives smaller errors, which would be consistent with that method's approach to dealing with noise.  A more detailed comparison is only warranted after further work has been put into optimizing the \ilium\ algorithm, such as the convergence for noisy data or the update clipping procedures.

SVM has the advantage of being much faster to apply once it has been tuned. As the speed of \ilium\ is probably dominated by the nearest neighbour initialization, this could be replaced with an SVM.  On the other hand, \ilium\ provides AP uncertainty estimates, goodness-of-fit estimates (for outlier/poor solution detection) and allows one to inspect the relevance of each input in determining the output for every object (equation~\ref{eqn:apcont}).

\end{document}